\documentclass[11pt]{article}
\pdfoutput=1
\usepackage[utf8x]{inputenc}	%for german Umlaute
\usepackage[english]{babel}
\usepackage{amssymb}
\usepackage{amsthm}
\usepackage{MnSymbol}
\usepackage{mathtools}		%for definition ":="
\usepackage{dsfont}		%for math modes
\usepackage{stmaryrd}		%GIT quotient
\usepackage{cite}		%for better citing
\usepackage[svgnames]{xcolor}
\usepackage{enumerate}
\usepackage{setspace}
\usepackage{multirow}
\usepackage{booktabs}
\usepackage{tikz}
\usetikzlibrary{shapes,snakes}
\usetikzlibrary{positioning}
\usetikzlibrary{chains}
\usetikzlibrary{arrows,fit,decorations.pathreplacing, shapes.misc}
\tikzstyle{every picture}+=[remember picture]
\tikzstyle{na} = [baseline=-.5ex]
\usepackage{xargs}
\usetikzlibrary{arrows.meta}
\usepackage{paralist}
\usepackage[font={small,it},labelfont=bf]{caption}
\usepackage{subcaption}
\usepackage[pdftex,breaklinks,colorlinks=true,urlcolor=RoyalBlue,linkcolor=blue,
citecolor=blue]{hyperref}		%for links on refs etc.
\catcode`@=11
\renewcommand{\section}{\@startsection{section}{1}{0pt}{\medskipamount}
{\medskipamount}{\Large\bf}}
\numberwithin{equation}{section}
\catcode`@=12

\newcommand{\im}{\mathrm{i}}

\def\rank{\mathrm{rk}}

\def\dim{\mathrm{dim}}

\newcommand{\HS}{\mathrm{HS}}
\newcommand{\HWG}{\mathrm{HWG}}

\newcommand{\PE}{\mathrm{PE}}

\newcommand{\C}{\mathbb{C}}
\newcommand{\R}{\mathbb{R}}

\newcommand{\HH}{\mathbb{H}}
\newcommand{\NN}{\mathbb{N}}  
\newcommand{\Z}{\mathbb{Z}}

\newcommand{\MCoulomb}{\mathcal{M}_C}
\newcommand{\Coulomb}{\mathcal{C}}
\newcommand{\MHiggs}{\mathcal{M}_H}
\newcommand{\Higgs}{\mathcal{H}}

\newcommand{\orbit}[1]{\mathcal{O}_{#1}}
\newcommand{\clorbit}[1]{\overline{\mathcal{O}}_{#1}}
\newcommand{\height}[1]{\text{ht}(#1)}

\newcommand{\stab}[2]{\mathrm{Stab}_{#1}(#2)}
\newcommand{\Ncal}{\mathcal{N}}

\newcommand{\gfrak}{\mathfrak{g}}  

\newcommand{\tfrak}{\mathfrak{t}}
  
\newcommand{\Wcal}{\mathcal{W}}
\newcommand{\uo}{{ U(1)}}

\newcommand{\glrm}{{{\rm GL}}}

\newcommand{\urm}{{{\rm U}}}

\newcommand{\surm}{{{\rm SU}}}
\newcommand{\surmL}{{{\mathfrak{su}}}}
\newcommand{\sorm}{{{\rm SO}}}
\newcommand{\orm}{{{\rm O}}}
\newcommand{\sormL}{{{\mathfrak{so}}}}

\newcommand{\slrmL}{{{\mathfrak{sl}}}}
\newcommand{\sprm}{{{\rm Sp}}}
\newcommand{\sprmL}{{{\mathfrak{sp}}}}

\newcommand{\utwo}{{{\rm U}(2)}}

\newcommand{\G}{\mathcal{G}}

\newcommand{\GNOgauge}{\widehat{\G}}
\newcommand{\GNOG}{\widehat{G}}

\newcommand{\Lie}{\mathrm{Lie}}

\newcommand{\Fcal}{\mathcal{F}}

\newcommand{\Rcal}{\mathcal{R}}
\newcommand{\Pcal}{\mathcal{P}}

\newcommand{\Ucal}{\mathcal{U}}

\newlength{\myline}% line thickness
\newcommandx*{\doublearrow}[4][1=0, 2=1]{
  \draw[line width=\myline,double distance=3\myline,#3] #4;
}

\newcommandx*{\triplearrow}[4][1=0, 2=1]{
  \draw[line width=\myline,double distance=5\myline,#3] #4;
  \draw[line width=\myline,shorten <=#1\myline,shorten >=#2\myline,#3] #4;
}
\allowdisplaybreaks
\begin{document}
\begin{titlepage}
\setcounter{page}{0}
\begin{flushright}
Imperial/TP/18/AH/04\\
UWTHPH-2018-15 
\end{flushright}

\vskip 2cm

\begin{center}

{\Large\bf Resolutions of nilpotent orbit closures \\
via Coulomb branches of $3$-dimensional $\Ncal=4$ theories
}

\vspace{15mm}

{\large Amihay Hanany${}^{1}$} , \ {\large Marcus Sperling${}^{2}$} 
\\[5mm]
\noindent ${}^1${\em Theoretical Physics Group, Imperial College London\\
Prince Consort Road, London, SW7 2AZ, UK}\\
{Email: {\tt a.hanany@imperial.ac.uk}}
\\[5mm]
\noindent ${}^{2}${\em Fakultät für Physik, Universität Wien}\\
{\em Boltzmanngasse 5, 1090 Wien, Austria}\\
Email: {\tt marcus.sperling@univie.ac.at}
\\[5mm]

\vspace{15mm}

\begin{abstract}
The Coulomb branches of certain $3$-dimensional $\Ncal=4$ quiver gauge theories 
are closures of nilpotent orbits of classical or exceptional Lie algebras. 
The monopole formula, as Hilbert series of the associated Coulomb branch 
chiral ring, has been successful in describing the singular hyper-Kähler 
structure.
By means of the monopole formula with background charges for flavour 
symmetries, which realises real mass deformations, we study the resolution 
properties of all (characteristic) height two nilpotent orbits. 
As a result, the monopole formula correctly reproduces (i) the existence of a 
symplectic resolution, (ii) the form of the symplectic resolution, and (iii) the 
Mukai flops in the case of multiple resolutions. 
Moreover, the (characteristic) height two nilpotent orbit closures are resolved 
by cotangent bundles of Hermitian symmetric spaces and the unitary Coulomb 
branch quiver realisations exhaust all the possibilities.
\end{abstract}

\end{center}

\end{titlepage}

{\baselineskip=12pt
{\footnotesize
\tableofcontents
}
}

%%%%%%%%%%%%%%%%%%%%%%%%%%%%%%%%%%%%%%%%%%%%%%%%%%%%%%%%%%%%%%%%%%%%%%%%%%%%%%%%
  \section{Introduction}
\label{sec:introduction}
Nilpotent orbits play an important role for supersymmetric gauge theories and 
appear whenever an embedding of $\surm(2)$ into some group is involved. In 
particular, nilpotent orbit closures have become the prototypical example of 
non-trivial hyper-Kähler singularities which can be realised as Coulomb and 
Higgs branches of $3$-dimensional $\Ncal=4$ gauge theories. This prominent role 
of nilpotent orbit closure is due to a theorem by Namikawa \cite{Namikawa:2016}.

The realisation that Coulomb $\MCoulomb$ and Higgs $\MHiggs$ branches can 
capture diverse features of the geometry of nilpotent orbits and Slodowy slices 
was due to the consideration of boundary conditions in $\Ncal=4$ supersymmetric 
gauge theories \cite{Gaiotto:2008ak}. 
The Higgs and Coulomb branch realisations of nilpotent orbits have been 
systematically developed in \cite{Hanany:2016gbz,Hanany:2017ooe}.
Recently \cite{Cabrera:2016vvv,Cabrera:2017njm}, another geometrical phenomenon 
has been realised in the Type IIB superstring set-up of $3$-dimensional 
$\Ncal=4$ theories: the classification of transverse slices and their 
singularities can be engineered by the \emph{Kraft--Procesi} transition, which 
is nothing else than a Higgs mechanism. Here, the partial 
ordering of nilpotent orbit closures and the notion of transverse slice allow to 
study the singularity structures of $\MCoulomb$ and $\MHiggs$. Uplifting this 
procedure to generic $3$-dimensional $\Ncal=4$ quiver gauge theories has turned 
out to be fruitful and is called \emph{quiver subtraction} 
\cite{Cabrera:2018ann}.

We take the recent advances as motivation to investigate the resolutions of the 
Coulomb branches which correspond to height two nilpotent orbit closures of 
classical and exceptional Lie algebras. 
The Coulomb branch geometry is affected by two types of deformations: complex 
and real mass deformations. 
Complex mass deformations are known to lead to deformations of $\MCoulomb$ and, 
interestingly, exhibit another part of the geometry of nilpotent orbits: they 
are geometrical incarnations of \emph{induced nilpotent orbits} 
\cite{Chacaltana:2012zy}. 
Moreover, complex mass deformations are accommodated for in the 
\emph{abelianisation} approach to $\MCoulomb$ of \cite{Bullimore:2015lsa}. 
However, this approach is insensitive to real mass deformations. In contrast, 
the Hilbert series of the Coulomb branch \cite{Cremonesi:2013lqa} is sensitive 
to real mass deformations, but not to complex masses. It is expected that
real mass deformations (at least partially) resolve the Coulomb 
branch singularities. 

As pioneered by \cite{Benvenuti:2006qr,Feng:2007ur,Cremonesi:2013lqa}, 
$\MCoulomb$ and $\MHiggs$ can be studied as algebraic varieties via the Hilbert 
series, which counts gauge invariant chiral operators on the associated chiral 
ring\footnote{We would like to emphasise that there is no need to (being able  
to) pick an entire sub algebra of the supersymmetry algebra; it is enough to 
pick a complex linear combination of supercharges to define the notion of 
chirality. This is crucial for $5$ and $6$ dimensional theories with $8$ 
supercharges.}.
The corresponding Hilbert series is called 
\emph{monopole formula} \cite{Cremonesi:2013lqa} because the relevant operators 
to consider on the chiral ring are monopole operators. 
The prescription of the monopole formula is capable to accommodate discrete 
real mass 
parameters via background charges (fluxes) for the flavour symmetry 
\cite{Cremonesi:2014kwa}; hence, one can 
utilise the Coulomb branch Hilbert series to study resolutions of the conical 
singularities.

The idea of introducing background charges into the monopole formula is not 
new, as it found applications in gluing techniques 
\cite{Cremonesi:2014kwa,Cremonesi:2014vla} or have been shown to be exchanged 
with baryonic background charges in the Hilbert series 
\cite{Forcella:2007wk,Butti:2007jv} of $\MHiggs$ upon mirror 
symmetry in \cite{Cremonesi:2014uva}. However, to the best of our knowledge, it 
has not been used to study resolutions of Coulomb branches systematically. 
Explicit examples include $3$-dimensional $\Ncal=4$ SQED in 
\cite[Section 2.5.2]{Cremonesi:2016nbo} and the study \cite{Hanany:2016djz} of 
cotangent bundles $T^*(G\slash H)$ of Kähler cosets, which are known to appear 
in the Springer resolution of nilpotent orbit closures. The latter approached 
the Hilbert series of $T^*(G\slash H)$ neither from the monopole formula nor 
with background charges present.
Moreover, resolutions of classical nilpotent orbit closures, realised by the 
so called $T_\rho^\sigma(G)$ theories, have been discussed in 
\cite{Cremonesi:2014kwa} and the information has been employed to derive the 
general Hilbert series in terms of Hall--Littlewood polynomials.

The outline of the remainder is as follows: in Section \ref{sec:preliminaries} 
we recall the relevant concepts such as nilpotent orbits and their 
symplectic resolutions as well as $3$-dimensional $\Ncal=4$ gauge theories, and 
the monopole formula. Thereafter, we systematically consider all 
(characteristic) height two orbits of classical algebras in Sections 
\ref{sec:A-type}---\ref{sec:D-type} and for exceptional algebras in Section 
\ref{sec:E-type}. We conclude and summarise in Section \ref{sec:conclusions}. 
The conventions for the calculations are provided in Appendix 
\ref{app:conventions}.

%%%%%%%%%%%%%%%%%%%%%%%%%%%%%%%%%%%%%%%%%%%%%%%%%%%%%%%%%%%%%%%%%%%%%%%%%%%%%%%%
   \section{Preliminaries}
\label{sec:preliminaries}
\subsection{Nilpotent orbits}
As we are concerned with quiver gauge theories whose Coulomb branches are 
closures of nilpotent orbits, we review the necessary ingredients. A general 
reference for nilpotent orbits is \cite{Collingwood:1993}.

Let $\gfrak$ be a semi-simple complex Lie algebra and $G$ its adjoint group.
We recall that nilpotent orbits $\orbit{}\subset \gfrak$, as complex adjoint 
orbits of $G$, are equipped with a canonical holomorphic symplectic from, the 
\emph{Kirillov-Kostant-Souriau form}. The fact that the adjoint orbits are 
hyper-Kähler varieties has been proven by Kronheimer 
\cite{Kronheimer:1990a,Kronheimer:1990b} in important special cases and by 
Biquard \cite{Biquard:1996} and Kovalev \cite{Kovalev:1996} in full generality.

The set of nilpotent orbits of $\gfrak$ is finite and the set 
of orbit closures admits a partial ordering via inclusion. Besides the trivial 
orbit, there exists a unique minimal orbit $\orbit{\text{min}}$, whose closure 
is contained in the closure of any other non-trivial nilpotent orbit. In 
addition, there exists a unique maximal orbit $\orbit{\text{max}}$, whose 
closure contains any other orbit closure. The closure $\clorbit{\text{max}}$ is 
called the nilpotent cone. Due to the nilpotency condition, any 
nilpotent orbit $\orbit{}$ is invariant under the dilation action of 
$\C^\times$ on $\gfrak$.
\paragraph{Classical algebras.}
Nilpotent orbits for the classical Lie algebras $\slrmL(n)$, $\sprmL(n)$, and 
$\sormL(n)$, can be labelled by partitions 
$\rho=(d_1,\ldots,d_t)$ of some $N\in \NN$, with $d_1\geq d_2 \geq \ldots \geq 
d_t$ and $\sum_{i=1}^t d_i = N$. Let $\Ncal(\gfrak)$ be the finite set of 
nilpotent 
orbits of $\gfrak$, then the following holds \cite[Proposition 
2.1]{Namikawa:2006}:
\begin{compactenum}
 \item[($\boldsymbol{A_n}$)] If $\gfrak=\surmL(n+1)$, then there exists a 
bijection between $\Ncal(\gfrak)$ and the set of partitions $\rho$ of $n+1$.
 \item[($\boldsymbol{B_n}$)] If $\gfrak=\sormL(2n+1)$, then there exists a 
bijection between $\Ncal(\gfrak)$ and the set of partitions $\rho$ of $2n+1$ 
such that 
even parts have even multiplicity.
\item[($\boldsymbol{C_n}$)] If $\gfrak=\sprmL(2n)$, then there exists a 
bijection between $\Ncal(\gfrak)$ and the set of partitions $\rho$ of $2n$ such 
that 
odd parts have even multiplicity.
\item[($\boldsymbol{D_n}$)] If $\gfrak=\sormL(2n)$, then there exists a 
surjection $f$ from $\Ncal(\gfrak)$ to the set of partitions $\rho$ of $2n$ 
such that even parts have even multiplicity. A partition of 
even parts only is called very even. For $\rho$ not a very even 
partition, $f^{-1}(\rho)$ consists of exactly one orbit. For $\rho$ very even, 
$f^{-1}(\rho)$ consists of exactly two different orbits.
\end{compactenum}
Alternatively, nilpotent orbits may be labelled by \emph{weighted Dynkin 
diagrams}, which are briefly discussed in appendix \ref{app:weighted_Dynkin}.
Next, we recall the \emph{height} $\height{\orbit{\rho}}$ of $\orbit{\rho}$ 
from 
\cite{Panyushev:1999}:
\begin{compactenum}[(i)]
 \item $\gfrak =\slrmL(n)$ or $\sprmL(n)$ then $\height{\orbit{\rho}}= 
2(d_1-1)$ 
  \item $\gfrak = \sormL(n)$ then  $\height{\orbit{\rho}}= 
  \begin{cases}
   d_1 +d_2 -2, & d_2 \geq d_1 -1 \\
   2d_1 -4, & d_2 \leq d_1 -2
  \end{cases}$  
\end{compactenum}
In this work, we restrict to nilpotent orbits of $\height{\orbit{}}=2$ for 
the following two reasons: (i) the closure of a nilpotent orbit of height 
$\height{\orbit{}}\leq 2$ always admits a unitary Coulomb branch quiver 
realisation \footnote{See a comment in \cite[Footnote 11]{Cabrera:2018ann}.} 
and (ii) the observation from \cite{Hanany:2016gbz,Hanany:2017ooe} 
that height two orbit closures have a simple Hilbert series or Highest Weight 
Generating function. The details will become clear in the subsequent sections.
\paragraph{Exceptional algebras.}
The classification of nilpotent orbits for exceptional algebras is more 
involved. One obvious obstacle to overcome is that exceptional groups do not 
act as matrices on their fundamental vector space. Several labelling methods 
for nilpotent orbits have been developed by Dynkin \cite{Dynkin:1957um}, 
Bala-Carter \cite{BalaCarter:1,BalaCarter:2}, and Hesselink 
\cite{Hesselink:1978}, to name a few.
Here, with the aim to study Coulomb branches, we follow the labelling by 
Characteristics and refer for details to 
\cite{Hanany:2017ooe}. The height of exceptional nilpotent orbits can be 
calculated following \cite[Section 2]{Panyushev:1999}; the definition agrees 
for classical algebras with the partition data given above.
% 
%%%%%%%%%%%%%%%%%%%%%%%%%%%%%%%%%%%%%%%%%%%%%%%%%%%%%%%
%
\subsection{Resolutions}
It is well-known that the closure of nilpotent orbit $\orbit{}$ is a singular 
(in general non-normal) variety.
By Hironaka's work \cite{Hironaka:1964}, any complex variety admits a 
resolution, but there may exist many different resolutions. One would like to 
restrict to certain ``good'' resolutions. For a symplectic variety, such a 
preferred resolution has been introduced by Beauville \cite{Beauville:2000}, 
denoted as \emph{symplectic resolution}. Roughly, for a symplectic variety $X$ 
and $\pi: Z \to X$ a resolution, then $\pi$ is a symplectic resolution if for 
any symplectic form $\omega$ on the regular part of $X$, the pull-back 
$\pi^*(\omega)$ extends to a symplectic form on $Z$.

For nilpotent orbits, it was proven by Panyushev \cite{Panyushev:1991} that the 
natural symplectic 2-form on $\orbit{}$ extends to any resolution of 
$\clorbit{}$, i.e.\ $\clorbit{}$ is a variety with a symplectic singularity. 
Subsequently, Fu \cite{Fu:2003a} determined all nilpotent orbit closures which 
admit a symplectic resolution.

One finds that every nilpotent orbit closure in $\slrmL(n+1)$ admits a 
symplectic resolution \cite[Proposition 5.1]{Fu:2006}. However, if one hopes 
that 
every nilpotent orbit closure admits a resolution, one finds the following 
disillusioning result \cite[Proposition 5.2]{Fu:2006}: For a simple Lie algebra 
$\gfrak$, the closure $\clorbit{\text{min}}$ admits a symplectic resolution if 
and only if $\gfrak$ is of $A$-type. The general statement \cite{Fu:2003a} is 
that a symplectic variety $\clorbit{}$ admits a symplectic resolution if and 
only if $\orbit{}$ is a \emph{Richardson orbit}, see \cite{Collingwood:1993} for 
a definition.
For the orbits which do admit a symplectic resolution, it is of the form 
\cite{Fu:2003a}
\begin{align}
 \pi : Z \to \clorbit{} \qquad \text{with} \quad Z \cong T^*\left( G \slash P 
\right)
\label{eq:resolution}
\end{align}
where $P\subset G$ is some parabolic subgroup --- called 
\emph{polarisation} of $\orbit{}$, see\cite{Hesselink:1978}.

Even when restricting the attention to symplectic resolutions $T^*(G\slash P) 
\to \clorbit{}$, there can exist several polarisations which yield different 
symplectic resolutions $T^*(G \slash P_i) \to \clorbit{}$. The 
rational map $\phi: T^*(G \slash P_1) \dashedrightarrow T^*(G \slash P_2)$ 
between any two resolutions is 
called a locally trivial family of \emph{Mukai flops}. Following 
\cite{Namikawa:2006,Fu:2007}, 
there are three basic types of Mukai flops: $A$, $D$, and $E_{6,I}$/ 
$E_{6,II}$. For a generic $\clorbit{}$, the birational map $\phi$ decomposes 
into a finite number of diagrams $Y_l \rightarrow X_l \leftarrow Y_{l+1}$, 
$l=1,\ldots,m$, with $Y_1=T^*(G \slash P_1)$ and $Y_m = T^*(G \slash P_2)$ such 
that each diagram is a basic Mukai flop. For closures of height two nilpotent 
orbits the basic Mukai flops suffice.
\begin{table}[t]
\centering
 \begin{tabular}{c|c|c}
 \raisebox{-.5\height}{
    	\begin{tikzpicture}
    	\tikzset{node distance = 0.5cm}
	\tikzstyle{gauge} = [circle, draw,inner sep=2.5pt];
	\tikzstyle{dark} = [circle,draw,inner sep=2.5pt,fill=black];
	\node (g1) [gauge] {};
	\node (g2) [right of =g1] {$\ldots$};
	\node (g3) [dark,right of =g2,label=below:{$k$}] {};
	\node (g4) [right of =g3] {$\ldots$};
	\node (g5) [gauge,right of =g4] {};
	\draw (g1)--(g2) (g2)--(g3) (g3)--(g4) (g4)--(g5);
	\end{tikzpicture}
	}
   & $A_{n}$ with $2k+1\neq n$ &
   \raisebox{-.5\height}{
    	\begin{tikzpicture}
    	\tikzset{node distance = 0.5cm}
	\tikzstyle{gauge} = [circle, draw,inner sep=2.5pt];
	\tikzstyle{dark} = [circle,draw,inner sep=2.5pt,fill=black];
	\node (g1) [gauge] {};
	\node (g2) [right of =g1] {$\ldots$};
	\node (g3) [dark,right of =g2,label=below:{$n{+}1{-}k$}] {};
	\node (g4) [right of =g3] {$\ldots$};
	\node (g5) [gauge,right of =g4] {};
	\draw (g1)--(g2) (g2)--(g3) (g3)--(g4) (g4)--(g5);
	\end{tikzpicture}
	} \\
	\midrule
\raisebox{-.5\height}{
    	\begin{tikzpicture}
    	\tikzset{node distance = 0.5cm}
	\tikzstyle{gauge} = [circle, draw,inner sep=2.5pt];
	\tikzstyle{dark} = [circle,draw,inner sep=2.5pt,fill=black];
	\node (g3) [gauge] {};
	\node (g1) [dark,above left of=g3] {};
	\node (g2) [gauge,below left of=g3] {};
	\node (g4) [gauge,right of =g3] {};
	\node (g5) [right of =g4] {$\ldots$};
	\node (g6) [gauge,right of =g5] {};
	\draw (g1)--(g3) (g2)--(g3) (g3)--(g4) (g4)--(g5) (g5)--(g6);
	\end{tikzpicture}
	}
   & $D_{n}$ with $n=$ odd  &
   \raisebox{-.5\height}{
    	\begin{tikzpicture}
    	\tikzset{node distance = 0.5cm}
	\tikzstyle{gauge} = [circle, draw,inner sep=2.5pt];
	\tikzstyle{dark} = [circle,draw,inner sep=2.5pt,fill=black];
	\node (g3) [gauge] {};
	\node (g1) [gauge,above left of=g3] {};
	\node (g2) [dark,below left of=g3] {};
	\node (g4) [gauge,right of =g3] {};
	\node (g5) [right of =g4] {$\ldots$};
	\node (g6) [gauge,right of =g5] {};
	\draw (g1)--(g3) (g2)--(g3) (g3)--(g4) (g4)--(g5) (g5)--(g6);	
\end{tikzpicture}
	} \\
	\midrule
\raisebox{-.5\height}{
    	\begin{tikzpicture}
    	\tikzset{node distance = 0.5cm}
	\tikzstyle{gauge} = [circle, draw,inner sep=2.5pt];
	\tikzstyle{dark} = [circle,draw,inner sep=2.5pt,fill=black];
	\node (g1) [dark] {};
	\node (g2) [gauge,right of =g1] {};
	\node (g3) [gauge,right of =g2] {};
	\node (g4) [gauge,right of =g3] {};
	\node (g5) [gauge,right of =g4] {};
	\node (g6) [gauge,above of =g3] {};
	\draw (g1)--(g2) (g2)--(g3) (g3)--(g4) (g4)--(g5) (g3)--(g6);
	\end{tikzpicture}
	}
   & $E_{6,I}$  &
   \raisebox{-.5\height}{
    	\begin{tikzpicture}
    	\tikzset{node distance = 0.5cm}
	\tikzstyle{gauge} = [circle, draw,inner sep=2.5pt];
	\tikzstyle{dark} = [circle,draw,inner sep=2.5pt,fill=black];
	\node (g1) [gauge] {};
	\node (g2) [gauge,right of =g1] {};
	\node (g3) [gauge,right of =g2] {};
	\node (g4) [gauge,right of =g3] {};
	\node (g5) [dark,right of =g4] {};
	\node (g6) [gauge,above of =g3] {};
	\draw (g1)--(g2) (g2)--(g3) (g3)--(g4) (g4)--(g5) (g3)--(g6);
\end{tikzpicture}
	} \\
	\midrule
\raisebox{-.5\height}{
    	\begin{tikzpicture}
    	\tikzset{node distance = 0.5cm}
	\tikzstyle{gauge} = [circle, draw,inner sep=2.5pt];
	\tikzstyle{dark} = [circle,draw,inner sep=2.5pt,fill=black];
	\node (g1) [gauge] {};
	\node (g2) [dark,right of =g1] {};
	\node (g3) [gauge,right of =g2] {};
	\node (g4) [gauge,right of =g3] {};
	\node (g5) [gauge,right of =g4] {};
	\node (g6) [gauge,above of =g3] {};
	\draw (g1)--(g2) (g2)--(g3) (g3)--(g4) (g4)--(g5) (g3)--(g6);
	\end{tikzpicture}
	}
   & $E_{6,II}$  &
   \raisebox{-.5\height}{
    	\begin{tikzpicture}
    	\tikzset{node distance = 0.5cm}
	\tikzstyle{gauge} = [circle, draw,inner sep=2.5pt];
	\tikzstyle{dark} = [circle,draw,inner sep=2.5pt,fill=black];
	\node (g1) [gauge] {};
	\node (g2) [gauge,right of =g1] {};
	\node (g3) [gauge,right of =g2] {};
	\node (g4) [dark,right of =g3] {};
	\node (g5) [gauge,right of =g4] {};
	\node (g6) [gauge,above of =g3] {};
	\draw (g1)--(g2) (g2)--(g3) (g3)--(g4) (g4)--(g5) (g3)--(g6);
\end{tikzpicture}
	} 
 \end{tabular} 
 \caption{List of dual marked Dynkin diagrams for the dual parabolic subgroups 
leading to the Mukai flops of type $A$, $D$, $E_{6,I}$, and $E_{6,I}$.}
\label{tab:Mukai_Dynkin}
\end{table}
\paragraph{Mukai flop of type $\boldsymbol{A}$.}
Let $x \in \surmL(n+1)$ be a nilpotent element of partition $(2^k,1^{n+1-2k})$ 
and $x \in 
\orbit{}$. Then there exist two polarisations $P$ and $P'$ of $x$ such that 
$P= \mathrm{S}(\urm(k)\times \urm(n+1-k))$  and $P'= 
\mathrm{S}(\urm(n+1-k) \times \urm(k) )$. Hence, 
$\clorbit{}$ admits two Springer resolutions 
\begin{align}
 T^* \left(\surm(n+1)\slash P \right) \xrightarrow{\; \pi \;} 
\clorbit{} 
\xleftarrow{\; \pi' \,} T^* 
\left(\surm(n+1)\slash P' \right) \; .
\label{eq:flop_A-type}
\end{align}
Note that the dual parabolic subgroups $P$, $P'$ can be read off from the 
marked Dynkin diagrams in Table \ref{tab:Mukai_Dynkin}.
If $2k<n+1$ then \eqref{eq:flop_A-type} is a flop \cite[Lemma 
3.1]{Namikawa:2006}; and if $2k=n+1$ then the two 
resolutions are isomorphic \cite[Remark 3.2]{Namikawa:2006}.
\paragraph{Mukai flops of type $\boldsymbol{D}$.}
Suppose $n=$ odd and $n\geq3$. Let $x\in \sormL(2n)$ be a nilpotent element of 
type $(2^{n-1},1^2)$ and $x\in \orbit{}$. Then there exist two choices of flags 
$P_+$ and $P_-$; hence, $\clorbit{}$ admits two Springer resolutions
\begin{align}
 T^* (\sorm(2n)\slash P_+) \xrightarrow{\; \pi_+ \;} \clorbit{} 
\xleftarrow{\; \pi_- \;} T^* 
(\sorm(2n)\slash P_-) \; .
\label{eq:flop_D-type}
\end{align}
From the marked Dynkin diagrams in Table \ref{tab:Mukai_Dynkin}, we read off 
the dual parabolic subgroups to be $P_\pm \cong \surm(n)\times \uo$. The $\pm$ 
subscript refers to the choice of the spinor node.
\paragraph{Mukai flops of type $\boldsymbol{E_6}$.}
The $E_{6,I}$ corresponds to the orbit with Bala-Carter label $2A_1$ or the 
Characteristic $\{1,0,0,0,1,0\}$.
The two dual parabolic subgroup $\cong \sorm(10)\times \sorm(2)$ can be read 
off from the marked Dynkin diagrams in Table \ref{tab:Mukai_Dynkin}.

The basic flop $E_{6,II}$ corresponds to the orbit with Bala-Carter label 
$A_2+2A_1$ or Characteristic $\{0,1,0,1,0,0\}$.
From the marked Dynkin diagrams in Table \ref{tab:Mukai_Dynkin}, we can read 
off the two dual parabolic subgroups $\cong \surm(5)\times \surm(2)\times \uo$.
\paragraph{Hermitian symmetric spaces.}
Since  (characteristic) height two nilpotent orbit closures are discussed 
below, it is useful to recall the Hermitian symmetric spaces 
(HSS).
 The HSS were first classified by Cartan \cite{Cartan:1935} and can be 
realised as homogeneous spaces $G \slash H$, see Table \ref{tab:HSS}. In 
terms of Cartan's classification of compact Riemannian symmetric spaces, the 
Hermitian symmetric spaces are the four infinite series $A_{III}$, $D_{III}$, 
$C_I$, $B_I \slash D_I$, and the two exceptional spaces $E_{III}$, $E_{VII}$.

As a remark,  since a HSS of the form $G\slash H$ is Hermitian as well as a 
symmetric homogeneous space, it follows that 
$G \slash H$ is also Kähler. Consequently, $T^* (G \slash H)$ is naturally 
hyper-Kähler and we will encounter the cotangent bundles of the HSS spaces 
below.

\begin{table}[t]
\centering
\begin{tabular}{c|c|c|c}
name & $G$ & $H$ & $ \dim_\C(G\slash H)$\\ \hline
$A_{III}$ & $\surm(n+m)$ & $S(\urm(n) \times \urm(m))$ & $ n\cdot m$ \\
$D_{III}$ & $\sorm(2n)$ & $\urm(n)$ & $\frac{1}{2} n(n-1) $ \\
$C_I$ & $\sprm(n)$ & $\urm(n)$ & $\frac{1}{2}n(n+1)$ \\
$B_{I} \slash D_{I}$ & $\sorm(n+2)$ & $\sorm(n)\times \uo$ & $n$  \\
$E_{III}$ & $E_6$ & $\sorm(10) \times \uo$ & $16$ \\
$E_{VII}$ & $E_7$ & $E_6 \times \uo$ & $27$
\end{tabular}
\caption{The four infinite series and the two exceptional cases of the 
Hermitian 
symmetric spaces.}
\label{tab:HSS}
\end{table}
% 
%%%%%%%%%%%%%%%%%%%%%%%%%%%%%%%%%%%%%%%%%%%%%%%
%%%%%%%%%%%%%%%%%%%%%%%%%%%%%%%%%%%%%%%%%%%%%%%
%
\subsection{3-dimensional \texorpdfstring{$\Ncal=4$}{N=4} gauge theories}
We consider a generic $3$-dimensional $\Ncal=4$ gauge theory with gauge group 
$\G$ and matter, in the form of hypermultiplets, transforming in some 
(quaternionic) representation $\oplus_I n_I \Rcal_I$ of $\G$. Depending on the 
multiplicities $n_I$ there exists a non-trivial flavour symmetry $G_F$, 
sometimes called ``Higgs branch'' global symmetry. In addition, if $\G$ 
contains abelian factors, there exists another global ``Coulomb branch'' 
symmetry $G_J$ which in the ultra-violet is given by $G_J^{\text{UV}} = \uo^{\# 
(\uo \text{ in } \G)}$, and may be enhanced in the infrared to a non-abelian 
group 
$G_J^{\text{IR}}$ whose maximal torus is at least\footnote{In the class of 
examples considered here, the ranks of UV and IR Coulomb branch global 
symmetry coincide.} $G_J^{\text{UV}}$.
In addition, there is a non-trivial R-symmetry group $\surm(2)_{\Coulomb} 
\times \surm(2)_{\Higgs}$, such that the three vector multiplet scalars are a 
triplet under $\surm(2)_\Coulomb$ and the hypermultiplets transform as doublets 
under $\surm(2)_\Higgs$.

In the absence of mass deformations, the vacuum moduli space has a rich 
structure as a union of several branches of the form $\cup_a \Coulomb_a \times 
\Higgs_a$. $\Coulomb_a$ is a hyper-Kähler space parametrised by vacuum 
expectation values (VEVs) of gauge-invariant combinations of vector multiplet 
scalars; whereas $\Higgs_a$ is a hyper-Kähler space parametrised by VEVs of 
gauge invariant combinations of the hypermultiplet scalars. The \emph{Coulomb 
branch} $\MCoulomb$ and \emph{Higgs branch} $\MHiggs$ arise as maximal branches 
with one factor being trivial.
The global symmetries $G_F$ and $G_J$ act on $\MHiggs$ and $\MCoulomb$, 
respectively. Moreover, these actions are associated to triplets of moment 
maps. Geometrically, $\MCoulomb$ and $\MHiggs$ are hyper-Kähler singularities 
with $\surm(2)_\Coulomb$ or $\surm(2)_\Higgs$ isometry, respectively.

The $3$-dimensional gauge theories with eight supercharges allow for two 
classes of deformation parameters: masses and FI 
parameters, which take values in a Cartan subalgebra of $G_F$ and $G_J$, 
respectively. Under the R-symmetry, the masses transform as triplet under 
$\surm(2)_\Coulomb$ and the FI parameters form a triplet under 
$\surm(2)_\Higgs$. It is a known feature that masses can deform and/or resolve 
the geometry of the Coulomb parts $\Coulomb_a$; while FI parameters deform / 
resolve the Higgs parts $\Higgs_a$. Restricting attention to the Coulomb 
branch, the triplet $\{m_i\}_{i=1}^3$ of masses decomposes into a \emph{complex 
mass} $m^\C=m_1+\im m_2$ and a \emph{real mass} $m^\R= m_3$. The two have 
different implications on the geometry: while the real mass leads to a 
(partial) resolution of the singularities of $\MCoulomb$, the complex mass will 
deform the geometry of $\MCoulomb$.

Starting from \cite{Gaiotto:2008ak}, it has been realised that nilpotent orbit 
closures appear as Coulomb and Higgs branches of $3$-dimensional $\Ncal=4$ 
gauge theories.
The relation between quiver graphs and nilpotent orbit closures has been 
established in the mathematics literature by at least 
\cite{Kraft:1979,Kobak:1996}. These mathematical construction are all on the 
Higgs branch.
Recently, Namikawa \cite{Namikawa:2016} proved the following: if 
all generators of a hyper-Kähler singularity with an $\surm(2)_R$ symmetry have 
spin $=1$ under $\surm(2)_R$, then the corresponding variety is a nilpotent 
orbit closure of the Lie algebra of its isometry group.
Consequently, nilpotent orbit closures are to be considered as the simplest 
non-trivial singular hyper-Kähler spaces.

In the following, we will consider quiver gauge theories with unitary gauge 
groups, which have enhanced non-abelian global symmetry $G_J^{\text{IR}}$ of 
$ABCDE$-type, in order to realise nilpotent orbit closures of 
$\Lie(G_J^{\text{IR}})$.  For quiver theories, one reads 
off the Dynkin diagram of the non-abelian part of $G_J^{\text{IR}}$ from the 
set of \emph{balanced nodes}. Recall, a gauge node is balanced if the number of 
flavours is equal to twice its  rank.
% 
%%%%%%%%%%%%%%%%%%%%%%%%%%%%%%%%%%%%%%%%%%%%%%%%%%%%%%%%%
%%%%%%%%%%%%%%%%%%%%%%%%%%%%%%%%%%%%%%%%%%%%%%%%%%%%%%%%%
%
\subsection{Coulomb branch realisations of nilpotent orbit closures}
In \cite{Gaiotto:2008ak} a class of $3$-dimensional superconformal field 
theories, denoted as $T_\sigma^\rho(G)$, has been introduced. These theories 
arise as infrared limits of linear quiver gauge theories with unitary or 
alternating orthogonal-symplectic gauge groups. Here, $G$ is considered as 
classical group with GNO dual $\GNOG$ \cite{Goddard:1976qe}; $\rho$ is a 
partition of $G$ and $\sigma$ 
is a partition of $\GNOG$, as defined above. By construction, the mirror 
of $T_\sigma^\rho(G)$ is $T_\rho^{\sigma}(\GNOG)$.
For classical $G$, all these theories can be seen as originating from Type IIB 
brane constructions.

In this work, we restrict ourself to the Coulomb branch of $T^\rho(G)$ 
theories, which are obtained from $T_\sigma^\rho(G)$ via $\sigma=(1,\ldots,1)$.
It has been established that the Coulomb branch of $T^\rho(G)$, which is 
equivalent to the Higgs branch of $T_\rho(\GNOG)$ as an algebraic variety, is a 
nilpotent orbit closure
\begin{align}
 \MCoulomb(T^\rho(G)) \cong \MHiggs(T_\rho(\GNOG)) \cong \clorbit{\rho^\vee} \,.
\end{align}
Here, we need a map $\vee: \rho \mapsto \rho^\vee$ that takes partitions of $G$ 
to partitions of the GNO dual $\GNOG$. Such a map is known 
\cite{Spaltenstein:1982,Barbasch:1985,Chacaltana:2012zy,Balasubramanian:2014jca} 
and is named \emph{Barbasch--Vogan} map, see also 
\cite[Sections 4.3--4.4]{Cabrera:2017njm}. For 
$G=\surm(n)$, one simply has $\rho^\vee= \rho^T$; whereas the other classical 
groups have slightly more involved prescriptions. Since we will be dealing 
with unitary quiver realisations of the $BCD$-type $T^\rho(G)$, the details of 
the Barbasch-Vogan map are not utterly important and we refer to 
\cite{Chacaltana:2012zy,Cremonesi:2014uva} for explicit expositions.
\paragraph{A-type.}
For $A$-type nilpotent orbits, the Coulomb branch quivers (as well as the Higgs 
branch quivers) are well-behaved and exhaust all possible nilpotent orbits of 
type $A$ as their moduli space. In particular on the Coulomb branch side, the 
quiver for $A_n$ orbits have exactly $n$ unitary gauge nodes, which allows to 
compare not only the dimension of the Coulomb branch, but also the full 
refinement of the Hilbert series, which enables for the 
decomposition into irreducible representations of $\surm(n+1)$. 
\paragraph{BCD-type.}
The Higgs branch quivers for $BCD$-type nilpotent orbit closures are built from 
alternating orthogonal-symplectic gauge nodes and their constructions exhausts 
all possible nilpotent orbits. However, the Coulomb branch constructions are 
problematic for a number of reasons. Many of the theories with orthogonal and 
symplectic gauge nodes are \emph{bad} in the sense of \cite{Gaiotto:2008ak}. In 
other words, the monopole formula defined by Lagrangian data is ill-defined and 
divergent. Computations of Coulomb branches with non-unitary gauge groups yield 
the correct unrefined Hilbert series of the claimed orbit closure, 
but fail to reproduce the fully refined Hilbert series. 
Fortunately, a unitary quiver construction for near to minimal $BCD$-type 
nilpotent orbit closures (of characteristic height two) has been presented in 
\cite{Hanany:2016gbz} by means of 
flavoured finite-type Dynkin diagrams.  We will 
focus on these unitary realisations for $BCD$-type nilpotent orbits.

The quiver gauge theories for $D_n$ have first been computed in 
\cite{Hanany:1999sj}; the same quivers appear in \cite{Henderson:2014} in the 
study of Slodowy slices of $B$ and $D$-type in the vicinity of the maximal 
(regular) nilpotent orbit of the respective algebra.
\paragraph{Exceptional types.}
It is notoriously difficult to obtain exceptional nilpotent orbit 
closures from standard gauge or string theory constructions. Higgs branch 
constructions are not available simply because exceptional groups do not act 
via matrices on a fundamental vector space. While Coulomb branch constructions 
for minimal orbits are known 
\cite{Intriligator:1996ex,Cremonesi:2013lqa,Cremonesi:2014xha} for some time, 
constructions for near to minimal nilpotent orbit closures have only been 
proposed very recently in \cite{Hanany:2017ooe}. This unitary quiver 
construction again employs flavoured Dynkin diagrams and is 
limited to the lower dimensional orbits.
\paragraph{Hilbert series}
The Hilbert series for the $T_\rho(G)$ theories have been presented in 
\cite{Cremonesi:2014kwa,Cremonesi:2014vla}; while the Hilbert series of the 
general class $T_\sigma^\rho(G)$ has been studied in \cite{Cremonesi:2014uva}.
It is worthwhile noting that the closures of minimal nilpotent orbits 
correspond to reduced single instanton moduli spaces. The Hilbert series of 
these have first been computed in 
\cite{Benvenuti:2010pq} and many other constructions are known 
\cite{Cremonesi:2013lqa,Cremonesi:2014xha,Hanany:2015hxa}.
The Hilbert series and HWG for nilpotent orbit closures of classical and  
exceptional groups have been systematically studied in 
\cite{Hanany:2016gbz,Hanany:2017ooe}. 
Special attention to the distinction between $\sorm(N)$ and $\orm(N)$ gauge 
groups in Coulomb branch realisations of $\sormL(n)$ nilpotent orbit has been 
given in \cite{Cabrera:2017ucb}.

In view of the form of the resolutions \eqref{eq:resolution}, the 
HWG for $T^* (G\slash P)$ have been evaluated in \cite{Hanany:2016djz} and 
found to agree with the Coulomb branch nilpotent orbit results. 
% 
%%%%%%%%%%%%%%%%%%%%%%%%%%%%%%%%%%%%%%%%%%%%%%%%%%%%%%%%%
%%%%%%%%%%%%%%%%%%%%%%%%%%%%%%%%%%%%%%%%%%%%%%%%%%%%%%%%%
%
\subsection{Monopole formula with background charges}
To study the resolutions of Coulomb branches, one turns on non-trivial real 
mass parameters. As these transform in the adjoint of the flavour symmetry, the 
inclusion of real mass parameters can be realised via background charges 
(fluxes) in the monopole formula 
\cite{Cremonesi:2014kwa,Cremonesi:2015dja,Cremonesi:2016nbo}. 
Suppose we are 
given a $3$-dimensional $\Ncal=4$ gauge theory with gauge group $\G$. The GNO 
dual group is denoted by $\GNOgauge$, the Weyl group for $\G$ (and $\GNOgauge$) 
is $\Wcal$, and $\Phi^+$ denotes the set of positive roots $\alpha$ of 
$\Lie(\G)$. Moreover, the matter content transforms in a representation 
$\oplus_I n_I \Rcal_I$ of $\G$ and a representation $\Rcal_F$ of the flavour 
symmetry $G_F$. 
The weight vectors of $\Rcal_I$ are denoted by $\rho$, and the weights of 
$\Rcal_F$ by $\tilde{\rho}$. 

Associated to the gauge group $\G$ are (dynamical) bare monopole operators, 
which are uniquely labelled by lattice points in the GNO weight lattice 
$\Gamma_{\GNOgauge}$ up to gauge equivalence \cite{Borokhov:2002cg}. Similarly, 
there exist (background) 
monopole operators associated to the flavour symmetry group $G_F$, which are 
uniquely labelled by the GNO lattice of $G_F$ (again, up to equivalence).

The monopole charge $q \in \Gamma_{\GNOgauge}$ breaks the gauge symmetry via 
adjoint Higgs mechanism to $H_q = \stab{\G}{q} \subset \G$. The resulting 
residual gauge theory may admit further gauge invariant chiral operators 
that can take non-trivial vacuum expectation values, which are accounted for by
the Hilbert series $P_{\G}(t^2,m)$ of the residual gauge symmetry 
\cite{Cremonesi:2013lqa}. 
The combination of non-trivial monopole background and VEVs in the residual 
gauge symmetry leads to dressed monopole operators.

Thus, we are ready to recall the Coulomb branch Hilbert series\footnote{In 
order to work with an integer grading, we choose the R-charge of the bare BPS 
monopole operators to be counted by $t^2$ instead of $t$.} in the presence 
of background charges:
\begin{align}
 \HS_p(t^2,z)= \sum_{q \in \Gamma_{\GNOgauge} \slash \Wcal } P_{\G}(t^2,q) 
\cdot t^{2\Delta(q,p)} \cdot z^{J(q)} 
\label{eq:monopole_formula}
\end{align}
where
\begin{align}
 \Delta(q,p) = \frac{1}{2} 
 \sum_{I} \sum_{\rho_I \in \Rcal_I} \sum_{\tilde{\rho} \in \Rcal_{F}} 
 | \rho(q) + \tilde{\rho}(p) | - \sum_{\alpha \in \Phi^+ } |\alpha(q)|
\label{eq:conformal_dim}
\end{align}
is the conformal dimension of a monopole operator with charges $(q,p)$, see 
\cite{Borokhov:2002cg,Gaiotto:2008ak,Benna:2009xd,Bashkirov:2010kz}. 
The dressing factors are given by $P_{\G}(t^2,q) = 
\prod_{i=1}^{\rank{\G}}1\slash(1-t^{2d_i})$ with $d_i$ are degrees of the 
Casimir 
invariants of $H_q$, see \cite{Cremonesi:2013lqa}.
Note that we can restrict the background charge to  $p \in 
\Gamma_{\widehat{G}_F} \slash \Wcal_F$. In addition, we have chosen to account 
for
the topological symmetries $G_J^{\mathrm{UV}} = \uo^{\# (\uo \text{ in } \G)}$ 
by fugacities 
$z\equiv(z_i)$ and their charge $J(z)$.
The map $J$ is a linear projection map from the GNO weight lattice to the 
Cartan subalgebra of the flavour symmetry.
Similarly to the discussion of \cite[Equation (3.8)]{Cremonesi:2014kwa}, one 
can 
remove an extra overall topological $\uo$. For a set of flavour nodes labelled 
by $N_i$, the physical flavour symmetry is $(\prod_i \urm(N_i)) \slash \uo$ 
rather than $(\prod_i \urm(N_i)) $.
\paragraph{Hilbert series generating function.}
The monopole formula, with or without background charges, presents a 
computational challenge due to the step-wise linear behaviour of 
$\Delta(q,p)$. In earlier works \cite{Hanany:2016ezz,Hanany:2016pfm}, 
we have introduced a method to systematise and partly overcome these 
complications by restricting to the domains of linearity of $\Delta$. In 
the absence of background charges, this procedure naturally leads to affine 
monoids organised by a fan.

The inclusion of background charges $p$ leads to non-central 
hyperplanes 
\begin{align}
 H_{\rho, \tilde{\rho}}(p) = 
\left\{q \in \tfrak \;  \big|\;  \rho(q) + \tilde{\rho}(p) =0 \right\} 
\end{align}
and corresponding closed half-spaces  $H^\pm_{\rho, \tilde{\rho}}(p)$. 
Here, $\tfrak$ denotes a Cartan subalgebra of $\gfrak$. To resolve 
\eqref{eq:conformal_dim}, one would intersect all possible half-spaces. 
Contrary to the $p\equiv 0$ case, the intersection is not necessarily a 
polyhedral cone, but generically a polyhedron. Although there exists a 
mathematical notion for the Hilbert series for the intersection of a 
polyhedron with a lattice \cite{Bruns:2016}, we choose to circumvent the 
resulting problem by computing the generating function for 
\eqref{eq:monopole_formula}.
Considering
\begin{align}
 \Fcal(t^2,z;y) = \sum_{p \in \Gamma_{\widehat{G}_F} \slash 
\Wcal_F} y^{p} \cdot \HS_p(t^2,z) \; ,
\end{align}
we realise that all the techniques of \cite{Hanany:2016ezz,Hanany:2016pfm} are 
straightforwardly applicable. Hence, we use this approach to compute 
$\HS_p(t^2,z)$ from $\Fcal(t^2,z;y)$.
\paragraph{Highest weight generating function.}
Having computed $\HS_p(t^2,z)$, the result might not be too illuminating. 
Fortunately, we can project the Hilbert series onto the \emph{Highest Weight 
Generating function} (HWG) \cite{Hanany:2014dia}. To summarise the essentials, 
the considered unitary quiver theories all have an enhanced infrared global 
symmetry $G_J^{\mathrm{IR}}$ of type $ABCDE$, which is counted in the refined HS 
by fugacities $z_i$, $i=1,\ldots, r$ with $r=\rank(G_J^{\mathrm{IR}})$. One 
transforms the HS into a character expansion of 
$\gfrak_J=\Lie(G_J^{\mathrm{IR}})$, via mapping the $z_i$ to new fugacities 
$x_i$ 
using the Cartan matrix of $\gfrak_J$, see appendix \ref{app:conventions}. 
Hence,
\begin{align}
 \HS_p(t^2,z)= \sum_{n\in \NN} f_n(z_i;p)  t^{n} 
 \quad \xrightarrow[\, \text{matrix}\,]{\text{Cartan}} \quad
 \HS_p(t^2,x) = \sum_{n\in \NN } \tilde{f}_n(x_i;p)  t^{n}
 \label{eq:HS_root_to_weight}
\end{align}
and each $\tilde{f}_n(x_i;p)$ can be decomposed into a finite sum of 
$G_J^{\mathrm{IR}}$-characters $\chi_{[n_1, \ldots,n_r]}(x_i)$. Next, one 
replaces each 
character by a monomial in highest weight fugacities $\mu_i$, $i=1,\ldots,r$
\begin{align}
 \chi_{[n_1, \ldots,n_r]}(x_i) \quad \mapsto \quad
\prod_{i=1}^r \mu_i^{n_i} \qquad \text{with} \quad n_i \in\NN
\end{align}
such that the HS is transformed into a HWG
\begin{align}
 \HWG_p(t^2,\mu) = \sum_{k_1,\ldots,k_r,n\in \NN} g_{k_1,\ldots,k_r;n}(p) \ 
\mu_1^{k_1}\cdot \ldots \cdot \mu_r^{k_r} t^n \;.
\end{align}
For a detailed introduction and the necessary computational tools, we refer to 
\cite{Hanany:2014dia}.
It is an empirical observation 
from \cite{Hanany:2016gbz,Hanany:2017ooe} that the Hilbert series for nilpotent 
orbit closures of (characteristic) height two have a simple HWG. As we will see 
below, this is also true for the Coulomb branches in the presence of background 
charges.
%%%%%%%%%%%%%%%%%%%%%%%%%%%%%%%%%%%%%%%%%%%%%%%%%%%%%%%%%%%%%%%%%%%%%%%%%%%%%%%%
%
   \section{A-type}
\label{sec:A-type}
We start by considering the nilpotent orbits of $A$-type of 
height two realised as a Coulomb branch.
To be specific, consider orbits $\orbit{\rho}$ of $A_n$ with partitions 
$\rho=(2^k,1^{n+1-2k})$ for $2\leq 2k \leq n+1$, such that 
$\height{\orbit{\rho}}=2$. 
The relevant Coulomb branch quiver gauge theories have been known for some 
time \cite{Gaiotto:2008ak} and the 
HWG for the singular case have been computed in \cite{Hanany:2016gbz} to read
\begin{align}
 \HWG^{(2^k,1^{n+1-2k})} (t^2)=\PE\left[\sum_{i=1}^k \mu_i 
\mu_{n+1-i} t^{2i} \right] \; .
\label{eq:HWG_A-type_no_flux}
\end{align}
We computed the Hilbert series and HWG in the presence of background charges to 
study the resolutions of the (closures) of the $A_n$ nilpotent orbits for 
$n\leq 5$. 
Moreover, due to \cite[Corollary 3.16]{Fu:2003a} we know that all height two 
nilpotent orbits admit a symplectic resolution (with a suitable polarisation). 
Now, we need to compare this fact to the Coulomb branch computations, for 
which we summarise the results in
Table \ref{tab:results_A-type_examples}.

\begin{table}[!ht]
\centering
 \begin{tabular}{c|c|c|c}
 \toprule
  $\rho$ & $\dim_\C$ & quiver & HWG with flux \\ \midrule
  $(2)$  &$2$ &
 	\raisebox{-.5\height}{
 	\begin{tikzpicture}
	\tikzset{node distance = 0.5cm}
	\tikzstyle{gauge} = [circle, draw,inner sep=2.5pt];
	\tikzstyle{flavour} = [regular polygon,regular polygon sides=4,inner 
sep=2.5pt, draw];
	\node (g1) [gauge,label=below:{$1$}] {};
	\node (f1) [flavour,above of=g1,label=above:{$2$}] {};
	\draw (g1)--(f1);
	\end{tikzpicture}
	}
&$ 
x_1^{p_1+p_2} (\mu_1 t)^{p_1-p_2} \ \PE[\mu_1^2 t^2] $
  \\
  \midrule
%   
%%%%%%%%%%%%%%%%%%%%%%%%%%%%%%%%%%%%%%%%%%%%%%
%
  $(2, 1) $ & $4$ &
  \raisebox{-.5\height}{
 	\begin{tikzpicture}
	\tikzset{node distance = 0.5cm}
	\tikzstyle{gauge} = [circle, draw,inner sep=2.5pt];
	\tikzstyle{flavour} = [regular polygon,regular polygon sides=4,inner 
sep=2.5pt, draw];
	\node (g1) [gauge,label=below:{$1$}] {};
	\node (g2) [gauge, right of=g1,label=below:{$1$}] {};
	\node (f1) [flavour,above of=g1,label=above:{$1$}] {};
	\node (f2) [flavour,above of=g2,label=above:{$1$}] {};
	\draw (g1)--(g2)
			(g1)--(f1)
			(g2)--(f2)
		;
	\end{tikzpicture}
	}
& $\begin{cases} x_1^{p_1} x_2^{p_2} (\mu_2 t)^{p_1-p_2} \ \PE[\mu_1 \mu_2 
t^2] &,\, p_1 \geq p_2 \\
 x_1^{p_1} x_2^{p_2} (\mu_1 t)^{p_2-p_1} \ \PE[\mu_1 \mu_2 
t^2] &,\, p_1 \leq p_2
\end{cases} $ \\
  \midrule
% 
%%%%%%%%%%%%%%%%%%%%%%%%%%%%%%%%%%%%%%%%%%%%%%%%%%%%%%% 
% 
  $ (2 , 1^2)$  & $ 6$ & 
\raisebox{-.5\height}{
 	\begin{tikzpicture}
	\tikzset{node distance = 0.5cm}
	\tikzstyle{gauge} = [circle, draw,inner sep=2.5pt];
	\tikzstyle{flavour} = [regular polygon,regular polygon sides=4,inner 
sep=2.5pt, draw];
	\node (g1) [gauge,label=below:{$1$}] {};
	\node (g2) [gauge, right of=g1,label=below:{$1$}] {};
	\node (g3) [gauge, right of=g2,label=below:{$1$}] {};
	\node (f1) [flavour,above of=g1,label=above:{$1$}] {};
	\node (f3) [flavour,above of=g3,label=above:{$1$}] {};
	\draw (g1)--(g2)
			(g2)--(g3)
			(g1)--(f1)
			(g3)--(f3)
		;
	\end{tikzpicture}
	}
& $\begin{cases} x_1^{p_1} x_3^{p_3} (\mu_3 t)^{p_1-p_3} \ \PE[\mu_1 \mu_3 
t^2] & ,\; p_1 \geq p_3\\
 x_1^{p_1} x_3^{p_3} (\mu_1 t)^{p_3-p_1} \ \PE[\mu_1 \mu_3 
t^2] & ,\; p_1 \leq p_3
\end{cases}$
 \\
$ (2^2)$ & $8$ &
 \raisebox{-.5\height}{
 	\begin{tikzpicture}
	\tikzset{node distance = 0.5cm}
	\tikzstyle{gauge} = [circle, draw,inner sep=2.5pt];
	\tikzstyle{flavour} = [regular polygon,regular polygon sides=4,inner 
sep=2.5pt, draw];
	\node (g1) [gauge,label=below:{$1$}] {};
	\node (g2) [gauge, right of=g1,label=below:{$2$}] {};
	\node (g3) [gauge, right of=g2,label=below:{$1$}] {};
	\node (f2) [flavour,above of=g2,label=above:{$2$}] {};
	\draw (g1)--(g2)
			(g2)--(g3)
			(g2)--(f2)
		;
	\end{tikzpicture}
	}
& $x_2^{p_1+p_2} (\mu_2 t^2)^{p_1-p_2} \ \PE[\mu_1 \mu_3 t^2 + \mu_2^2 
t^4] $
 \\
  \midrule
%  
%%%%%%%%%%%%%%%%%%%%%%%%%%%%%%%%%%%%%%%%%%%% 
% 
  $( 2 , 1^3)$ & $8$&
\raisebox{-.5\height}{
 	\begin{tikzpicture}
	\tikzset{node distance = 0.5cm}
	\tikzstyle{gauge} = [circle, draw,inner sep=2.5pt];
	\tikzstyle{flavour} = [regular polygon,regular polygon sides=4,inner 
sep=2.5pt, draw];
	\node (g1) [gauge,label=below:{$1$}] {};
	\node (g2) [gauge, right of=g1,label=below:{$1$}] {};
	\node (g3) [gauge, right of=g2,label=below:{$1$}] {};
	\node (g4) [gauge, right of=g3,label=below:{$1$}] {};	
	\node (f1) [flavour,above of=g1,label=above:{$1$}] {};
	\node (f4) [flavour,above of=g4,label=above:{$1$}] {};
	\draw (g1)--(g2)
			(g2)--(g3)
			(g1)--(f1)
			(g4)--(f4)
			(g3)--(g4);
	\end{tikzpicture}
	}
& $ \begin{cases} x_1^{p_1} x_4^{p_4} (\mu_4 t)^{p_1-p_4} \ \PE[\mu_1 \mu_4 
t^2] &, \; p_1 \geq p_4\\
 x_1^{p_1} x_4^{p_4} (\mu_1 t)^{p_4-p_1} \ \PE[\mu_1 \mu_4 
t^2] & , \; p_1 \leq p_4 \end{cases}$
 \\
$( 2^2 , 1)$ &  $12 $ &
\raisebox{-.5\height}{
 	\begin{tikzpicture}
	\tikzset{node distance = 0.5cm}
	\tikzstyle{gauge} = [circle, draw,inner sep=2.5pt];
	\tikzstyle{flavour} = [regular polygon,regular polygon sides=4,inner 
sep=2.5pt, draw];
	\node (g1) [gauge,label=below:{$1$}] {};
	\node (g2) [gauge, right of=g1,label=below:{$2$}] {};
	\node (g3) [gauge, right of=g2,label=below:{$2$}] {};
	\node (g4) [gauge, right of=g3,label=below:{$1$}] {};	
	\node (f2) [flavour,above of=g2,label=above:{$1$}] {};
	\node (f3) [flavour,above of=g3,label=above:{$1$}] {};
	\draw (g1)--(g2)
			(g2)--(g3)
			(g2)--(f2)
			(g3)--(f3)
			(g3)--(g4);
	\end{tikzpicture}
	}
 & $ \begin{cases}
 x_2^{p_2} x_3^{p_3} (\mu_3 t^2)^{p_2-p_3} \ \PE[\mu_1 \mu_4 t^2 + \mu_2 
\mu_3 t^4] & ,\; p_2 \geq p_3 \\
x_2^{p_2} x_3^{p_3} (\mu_2 t^2)^{p_3-p_2} \ \PE[\mu_1 \mu_4 t^2 + \mu_2 
\mu_3 t^4] & ,\; p_2 \leq p_3
\end{cases}$
 \\
  \midrule
%   
%%%%%%%%%%%%%%%%%%%%%%%%%%%%%%%%%%%%%%%%%%%%%%%%%%%%%%%%%%%%%
%
  $( 2 , 1^4)$ & $10 $&
\raisebox{-.5\height}{
 	\begin{tikzpicture}
 	\tikzset{node distance = 0.5cm}
	\tikzstyle{gauge} = [circle, draw,inner sep=2.5pt];
	\tikzstyle{flavour} = [regular polygon,regular polygon sides=4,inner 
sep=2.5pt, draw];
	\node (g1) [gauge,label=below:{$1$}] {};
	\node (g2) [gauge, right of=g1,label=below:{$1$}] {};
	\node (g3) [gauge, right of=g2,label=below:{$1$}] {};
	\node (g4) [gauge, right of=g3,label=below:{$1$}] {};
	\node (g5) [gauge, right of=g4,label=below:{$1$}] {};	
	\node (f1) [flavour,above of=g1,label=above:{$1$}] {};
	\node (f5) [flavour,above of=g5,label=above:{$1$}] {};
	\draw (g1)--(g2)
			(g2)--(g3)
			(g1)--(f1)
			(g5)--(f5)
			(g3)--(g4)
			(g4)--(g5);
	\end{tikzpicture}
	}
& $ \begin{cases} x_1^{p_1} x_5^{p_5} (\mu_5 t)^{p_1-p_5} \ \PE[\mu_1 \mu_5 
t^2] &, \; p_1 \geq p_5\\
x_1^{p_1} x_5^{p_5} (\mu_1 t)^{p_5-p_1} \ \PE[\mu_1 \mu_5 
t^2] &, \; p_1 \leq p_5 
\end{cases}$
\\
  $( 2^2 , 1^2)$ & $16 $ &
\raisebox{-.5\height}{
 	\begin{tikzpicture}
	\tikzset{node distance = 0.5cm}
	\tikzstyle{gauge} = [circle, draw,inner sep=2.5pt];
	\tikzstyle{flavour} = [regular polygon,regular polygon sides=4,inner 
sep=2.5pt, draw];
	\node (g1) [gauge,label=below:{$1$}] {};
	\node (g2) [gauge, right of=g1,label=below:{$2$}] {};
	\node (g3) [gauge, right of=g2,label=below:{$2$}] {};
	\node (g4) [gauge, right of=g3,label=below:{$2$}] {};
	\node (g5) [gauge, right of=g4,label=below:{$1$}] {};	
	\node (f2) [flavour,above of=g2,label=above:{$1$}] {};
	\node (f4) [flavour,above of=g4,label=above:{$1$}] {};
	\draw (g1)--(g2)
			(g2)--(g3)
			(g2)--(f2)
			(g4)--(f4)
			(g3)--(g4)
			(g4)--(g5);
	\end{tikzpicture}
	  }
 & $ \begin{cases} x_2^{p_2} x_4^{p_4} (\mu_4 t^2)^{p_2-p_4} \ \PE[\mu_1 \mu_4 
t^2 + \mu_2 \mu_3 t^4] & , \; p_2 \geq p_4 \\
x_2^{p_2} x_4^{p_4} (\mu_2 t^2)^{p_4-p_2} \ \PE[\mu_1 \mu_4 t^2 + \mu_2 
\mu_3 t^4] &, \; p_2 \leq p_4 \end{cases} $
 \\
$(2^3)$ & $18 $ &
\raisebox{-.5\height}{
 	\begin{tikzpicture}
	\tikzset{node distance = 0.5cm}
	\tikzstyle{gauge} = [circle, draw,inner sep=2.5pt];
	\tikzstyle{flavour} = [regular polygon,regular polygon sides=4,inner 
sep=2.5pt, draw];
	\node (g1) [gauge,label=below:{$1$}] {};
	\node (g2) [gauge, right of=g1,label=below:{$2$}] {};
	\node (g3) [gauge, right of=g2,label=below:{$3$}] {};
	\node (g4) [gauge, right of=g3,label=below:{$2$}] {};
	\node (g5) [gauge, right of=g4,label=below:{$1$}] {};	
	\node (f3) [flavour,above of=g3,label=above:{$2$}] {};
	\draw (g1)--(g2)
			(g2)--(g3)
			(g3)--(f3)
			(g3)--(g4)
			(g4)--(g5);
	\end{tikzpicture}
	  }
	  &  
	  $x_3^{p_1+p_2} (\mu_3 t^3)^{p_1-p_2} \PE[\mu_1\mu_5 t^2 
+\mu_2\mu_4 t^4 + \mu_3^2 t^6]  $
 \\
 \bottomrule
 \end{tabular}
\caption{Coulomb branch quiver gauge theories for A-type algebras: gauge 
theories $T^{\rho^T}[\surm(n+1)]$ as Coulomb branch realisations of the 
(closures of the) nilpotent orbits $\clorbit{\rho}$ of $A_{n}$, for 
$n=1,2,3,4,5$. The unphysical $\uo$ in $G_F$ can be eliminated in the HWG by 
imposing that the sum of fluxes vanishes. In more detail, the $\uo$ could be 
counted by an auxiliary fugacity $z_0$ and one can impose $z_0^{p_k+p_{n+1-k}} 
\prod_{i=1}^n z_i^{r_i}=1$, where $r_i$ are the ranks of the $n$ gauge 
groups. Setting $z_0=1$ and converting root space to weight space fugacities, 
one obtains the condition $x_k x_{n+1-k}=1$.}
\label{tab:results_A-type_examples}
\end{table}

\paragraph{Minimal nilpotent orbit.}
Before considering the generic case, we elaborate on Coulomb branch quivers for 
the $A$-type minimal nilpotent closure for two reasons: firstly, to exemplify 
the calculations and explain the conclusions drawn from it. Secondly, to 
highlight the special geometry of these abelian theories, which have 
$\dim_\HH=n$ Coulomb branches with a $(\C^\times)^n$-action.

From the quiver representation with $n$ $\uo$ gauge nodes we read off the 
conformal dimension
\begin{align}
\raisebox{-.5\height}{
  	\begin{tikzpicture}
	\tikzstyle{gauge} = [circle, draw];
	\tikzstyle{flavour} = [regular polygon,regular polygon sides=4, draw];
	\node (g1) [gauge,label=below:{$1$}] {};
	\node (g2) [gauge, right of=g1,label=below:{$1$}] {};
	\node (g4) [right of=g2] {$\ldots$};
	\node (g3) [gauge, right of=g4,label=below:{$1$}] {};
	\node (f1) [flavour,above of=g1,label=above:{$1$}] {};
	\node (f3) [flavour,above of=g3,label=above:{$1$}] {};
	\draw (g1)--(g2)
			(g2)--(g4)
			(g4)--(g3)
			(g1)--(f1)
			(g3)--(f3)
		;
	\end{tikzpicture}
	}
	\qquad \rightarrow \quad 
	\Delta(q,p)= \frac{1}{2}\left( \sum_{i=1}^{n-1} |q_i -q_{i+1}| + |q_1 
-p_1| + 
|q_n -p_n| \right) \; ,
\end{align}
with magnetic charges $q_i\in \Z$, for $i=1,\ldots,n$, and fluxes $p_1,p_n\in 
\Z$.
Without loss of generality, we consider the case $p_1\geq p_N$.  
In the spirit of \cite{Hanany:2016ezz,Hanany:2016pfm}, we understand the 
absolute values in $\Delta(q,p)$ as defining hyperplanes in $\R^n$. Their 
intersection gives 
rise to bounded regions, i.e.\ polytopes, as well as unbounded regions, i.e.\ 
polyhedra. Since any polyhedron can be decomposed as Minkowski sum of a 
polytope and a polyhedral cone, we only consider the polytopes because we 
elaborated on how to deal with polyhedral cones in 
\cite{Hanany:2016ezz,Hanany:2016pfm}. 
Let us focus on the maximal dimensional polytope appearing in the summation 
range $q_i \in \Z$, $i=1,\ldots,n$.

Considering for a moment $\vec{q}\equiv (q_i)\in \R^n$, then the polytope 
$\Pcal_z$ is defined by the intersection of the following half-spaces
 \begin{align}
 P_z= \left\{\vec{q}\in \R^n \, | \, q_1-p_1\leq0 \right\} \cap 
\left\{\vec{q}\in \R^n \, | \, q_n -p_n \geq0 \right\}  \cap 
\bigcap_{i=1}^{n-1} \left\{\vec{q}\in \R^n \, | \, q_i - q_{i+1} \geq 0 
\right\} \subset \R^n
\end{align}
 and can equivalently be characterised by its vertices as follows:
 \begin{align}
 \Pcal_z= \mathrm{Conv} 
 \begin{Bmatrix}
  (p_n,p_n,\ldots , p_n, p_n)\\
  (p_1,p_n,\ldots, p_n, p_n)\\
  \ldots \\
  (p_1,p_1,\ldots , p_1, p_n) \\
  (p_1,p_1,\ldots , p_1, p_1) 
 \end{Bmatrix} \subset \R^n \, .
 \end{align}
From the refined monopole formula \eqref{eq:monopole_formula}, we see that 
$\Pcal_z$ is a polytope in the root lattice, spanned by $z_1, \ldots,z_n$.
Next, we utilise the Cartan matrix of $A_n$ to map $\Pcal_z$ via 
\eqref{eq:root_to_weight} into a polytope $\Pcal_x$ in the weight lattice, 
spanned by $x_1,\ldots, x_n$. This is exactly the same transformation as in 
\eqref{eq:HS_root_to_weight}. The polytope $\Pcal_x$ is again defined by its 
vertices:
\begin{align}
\Pcal_x= \mathrm{Conv}
\begin{Bmatrix}
 (p_n, 0,\ldots,0,p_n)  \\
 (2p_1 -p_n,p_n-p_1, \ldots, 0,p_n )  \\
  \ldots  \\
   (p_1,0,\ldots,p_1-p_n ,2p_n -p_1) \\
  (p_1, 0,\ldots, 0,p_1) 
\end{Bmatrix} \subset \R^n\; . 
\end{align}
 Splitting off an off-set vector $ (p_1, 0,\ldots,0,p_n)$ and realising 
a dilation factor $p_1 - p_n \geq0$, we rewrite the polytope $\Pcal_x$ as 
\begin{subequations}
\begin{align}
 \Pcal_x&=  ( p_1, 0,\ldots,0,p_n)\ + \ (p_1-p_n) \times \mathrm{Conv}(S) \\
S&=
\begin{Bmatrix}
 (-1, 0,\ldots,0,0)  \\
 (1,-1, \ldots, 0,0 )  \\
  \ldots  \\
   (0,0,\ldots,1 ,-1) \\
  (0, 0,\ldots, 0,1) \\
\end{Bmatrix} \; .
\end{align}
\end{subequations}
Taking the intersection with the GNO weight lattice $\Z^n$, we observe that 
$\Pcal_x \cap \Z^n$ agrees with the weight vectors of the $\surm(n+1)$ 
representation $[0,0,\ldots,0,p_1-p_n]$ shifted by an off-set $( p_1, 
0,\ldots,0,p_n)$. This follows because the set $S\cap \Z^n$ agrees with the 
weights of $[0,0,\ldots,0,1]$, and $p_1-p_n$ yields the 
dilation to $[0,0,\ldots,0,p_1-p_n]$.
Consequently, the contribution of $\Pcal_x \cap \Z^n$ to the HWG is
\begin{align}
 \HWG_{(p_1,p_n)}(\Pcal_x \cap \Z^n) = x_1^{p_1} x_n^{p_n} \ (\mu_n 
t)^{p_1-p_n} \,. \label{eq:min_A-type_polytope}
\end{align}
Comparing to the full HWG for $(2,1^{n-1})$, as shown in 
Table \ref{tab:results_A-type_examples} or below in \eqref{eq:HWG_A-type}, we 
see that \eqref{eq:min_A-type_polytope} describes the ratio of the HWG 
with and without background fluxes. The case $p_n \geq p_1$ produces the 
representation $[p_n-p_1,0,\ldots,0]$ instead, such that the HWG is changed 
appropriately.

\begin{figure}[t]
\centering
\begin{tikzpicture}
  \draw [thick, black] (0,0) -- (4,0);%
  \draw [thick, black] (4,0) -- (4,4);%
  \draw [thick, black] (0,0) -- (4,4);%
  \draw (-0.15,-0.25) node {$(p_2,p_2)$};
  \draw (4+0.15,-0.25) node {$(p_1,p_2)$};
  \draw (4+0.15,4+0.2) node {$(p_1,p_1)$};
  \draw [red,-latex] (0.5,0.15) -- (1,0.15);%
  \draw (0.75,0.35) node {$X$};%
  \draw [red,-latex] (3.5,0.15) -- (3,0.15);%
  \draw (3.25,0.35) node {$U$};%
  \draw [red,-latex] (4.25,3.5) -- (4.25,3);%
  \draw (4.5,3.25) node {$W$};%
  \draw [red,-latex] (4.25,0.5) -- (4.25,1);%
  \draw (4.5,0.75) node {$V$};%
  \draw [red,-latex] (0.25,0.75) -- (0.75,1.25);%
  \draw (0.4,1.3) node {$Z$};%
  \draw [red,-latex] (3.5,3.75) -- (3.0,3.25);%
  \draw (3.3,3.75) node {$Y$};%
\end{tikzpicture}%
\caption{The polytope $\Pcal_z$ arising in the monopole formula for the 
minimal nilpotent orbit closure of $A_2$. }
\label{fig:polytope_CP2}
\end{figure}
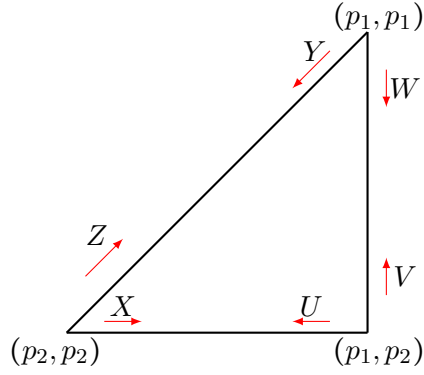

More geometrically, we recognise $S\cap \Z^n$ as standard simplex in $\R^n$. 
Since abelian $3$-dimensional $\Ncal=4$ theories are known to have hyper-toric 
Higgs and Coulomb branches, we might be tempted to take the standard simplex as 
indicator for a $\C P^n$. 
In fact, we would understand the $(p_1{
-}p_n)$-dilated simplex as giving the $T^* \C P^n$, where the flux $(p_1-p_n)$ 
determines the size of the $\C P^n$. Similarly to SQED with $N$ 
flavours \cite{Cremonesi:2014kwa,Cremonesi:2016nbo}, we can define operators on 
the vertices of $\Pcal_z\cap \Z^n$ which realise the correct transition 
functions between the affine patches of $\C P^n$. Let us illustrate this for 
$\C P^2$ as follows:
For $n=2$ the polyhedron $\Pcal_z$  is defined by the three edges 
$(p_1,p_1)$, $(p_1,p_2)$, and $(p_2,p_2)$. Define the following 
operators/coordinates, see also Figure \ref{fig:polytope_CP2}: 
\begin{subequations}
\begin{alignat}{2}
 &\Ucal_1 : \qquad &
V_{(p_2+a+b,p_2+b)} &= V_{(p_2,p_2)} X^a Z^b  \\
 &\Ucal_2 : \qquad &
 V_{(p_1-c,p_2+d)} &= V_{(p_1,p_2)} U^c V^d \\
&\Ucal_3 : \qquad &
 V_{(p_1-e,p_1-e-f)} &= V_{(p_1,p_1)} Y^e W^f 
\end{alignat}
\end{subequations}
On the overlap of, say, $\Ucal_1 \cap \Ucal_2$ we find
\begin{align}
 & & V_{(p_2 + a+b,p_2 +b)} 
 &= V_{(p_1 - (p_1-p_2 -a-b), p_2+b)} \quad \forall a,b \nonumber \\
&\Leftrightarrow \quad &
 V_{(p_2,p_2)} X^a Z^b 
 &= V_{(p_1,p_2)} U^{p_1-p_2 -a-b} V^b \quad \forall a,b \nonumber \\
&\Leftrightarrow \quad &
V_{(p_2,p_2)} (XU)^a (ZU)^b &= V_{(p_1,p_2)} U^{p_1-p_2 } V^b 
\quad \forall a,b  \nonumber\\
&\Leftrightarrow \quad  &
XU &= 1 \; , \qquad ZU= V \; ,
\end{align}
which are precisely the transition functions of $\C P^2$. 

In view of the known resolutions $\pi: T^* (\C P^n) \to \clorbit{\text{min}}$, 
it is suggestive to interpret the contribution of the bounded summation region 
as giving $T^*\C P^n$, where $\C P^n$ is of size $(p_1-p_n)$. 
The size can also be seen from the factor $t^{p_1 -p_n}$ in 
\eqref{eq:min_A-type_polytope}. This follows, because for 
all $p_1 -p_n >0$ the geometry of the resolved space is $T^*(\C P^n)$, but 
the size of the $\C P^n$ is not fixed yet.
The entire argument 
becomes even more compelling by recalling that the two resolutions for 
$p_1\geq p_n$ and $p_1\leq p_n$ are manifestations of the Mukai flop of type 
$A$, see \eqref{eq:flop_A-type}. 

\paragraph{General height two case.}
To begin with, we observe that all Coulomb branch quiver gauge 
theories allow for non-trivial resolution parameters, as the flavour 
symmetry groups are either $(\uo{\times} \uo)\slash \uo$ or $\surm(2)$.
From the examples computed, we can even conjecture the Coulomb 
branch Hilbert series 
in the presence of background charges for all $\clorbit{(2^k,1^{n+1-2k})}$ of 
$\slrmL(n+1)$. We propose
\begin{align}
 \HWG^{(2^k,1^{n+1-2k})}_{(p_k,p_{n+1-k})}(t^2)=
 \begin{cases}
 x_k^{p_k} x_{n+1-k}^{p_{n+1-k}}  \ 
 \left(\mu_{n+1-k} t^k \right)^{p_k-p_{n+1-k}} \
\PE\left[\sum_{i=1}^k \mu_i \mu_{n+1-i} t^{2i} \right] \, , &  p_k \geq 
p_{n+1-k} 
\\
x_k^{p_k} x_{n+1-k}^{p_{n+1-k}} \ 
\left(\mu_{k} t^k \right)^{p_{n+1-k}-p_k} \
\PE\left[\sum_{i=1}^k \mu_i \mu_{n+1-i} t^{2i} \right] \, , & p_k<p_{n+1-k}
\end{cases}\,.
\label{eq:HWG_A-type}
\end{align}
To remove the overall $\uo$ shift symmetry in $G_F$, one simply has to impose 
$p_k + p_{n+1-k} =0$. This reduces the problem to one effective resolution 
parameter $\sim \pm(p_k - p_{n+1-k})$.
Inspecting the expression \eqref{eq:HWG_A-type}, we observe two prominent 
features: firstly, the HWG of the resolved space factors into the HWG 
\eqref{eq:HWG_A-type_no_flux} of the singular space times a prefactor. Secondly, 
depending on the ordering of $p_k$ and $p_{n+1-k}$ the HWG becomes case 
dependent. Now, we aim to explain these observations.

As elaborated earlier, $\clorbit{\rho}$ is resolved via $\pi_\rho : T^*(G 
\slash P_\rho) \to \clorbit{\rho}$ with $\pi_\rho^{-1}(0) \cong G  \slash 
P_\rho$. For 
the height two partitions $\rho=(2^k,1^{n+1-2k})$, the relevant coset spaces 
are the Grassmann manifolds
\begin{align}
 G_{m,k}\cong \frac{\surm(m+k)}{\mathrm{S}(\urm(m)\times\urm(k))} \, ,
\end{align}
which enjoy the isomorphism $ G_{m,k} \cong G_{k,m}$. Moreover, for 
$k=1$ one obtains $G_{m,1} \cong \C P^m$. 
Consequently, the height two orbits of $A$-type are resolved by cotangent 
bundles of the Hermitian symmetric spaces $G_{m,k}$.

It is known that these spaces also appear as (semi-simple) coadjoint orbits of 
the fundamental weights $\mu_k$, $k=1,\ldots,n$ of $A_n$. To see this, note 
that the stabilisers are given by  
\begin{align}
 \stab{\surm(n+1)}{\mu_k} \cong S(\urm(k)\times \urm(n+1-k)) \;,
\end{align}
and observe that $ \stab{\surm(n+1)}{\mu_k} \cong  
\stab{\surm(n+1)}{\mu_{n+1-k}}$. Thus, we obtain the semi-simple orbits
\begin{align}
 \orbit{\mu_k}^{\mathrm{ss}}= G_{k,n+1-k} \cong G_{n+1-k,k} = 
\orbit{\mu_{n+1-k}}^{\mathrm{ss}} \,.
\end{align}
Hence, we identify the prefactor $\mu_k$ or $\mu_{n+1-k}$ in 
\eqref{eq:HWG_A-type} as accounting for the holomorphic sections on the 
cotangent bundle $T^* G_{k,n+1-k}$ 
over the exceptional fibre $\cong \orbit{\mu_k}^{\mathrm{ss}}= G_{k,n+1-k}$.

Next, the existence of two resolutions for $2k< n+1$ is the manifestation of 
the Mukai 
flop of type $A$. In other words, each of the height two nilpotent 
orbits has two symplectic resolutions, which are related by a Mukai flop 
\eqref{eq:flop_A-type}. The 
observation that the HWG changes depending on the relative sign of $p_k - 
p_{n+1-k}$ is consistent with this statement, as the prefactor change from 
$\mu_k$ to $\mu_{n+1-k}$ indicates the Mukai flop from $G_{k,n+1-k}$ to 
$G_{n+1-k,k}$. Consistently, we observe for $2k=n+1$, i.e.\ for the examples 
$\rho=(2)$ of $A_1$, $\rho=(2^2)$ of $A_3$, and $\rho=(2^3)$ of $A_5$, etc.\  
that 
only one resolution exists, because the two potential resolutions are isomorphic 
as discussed below \eqref{eq:flop_A-type}. (In this case, it is implicitly 
understood that the fluxes in \eqref{eq:HWG_A-type} are relabelled to 
$p_1$, $p_2$ satisfying $p_1\geq p_2$.)
\paragraph{Higgs branch.}
To complete the aforementioned reasoning, consider the mirror quiver of 
partition $\rho^T = (n+1-k,k)$: i.e.\ SQCD with $\urm(k)$ gauge group and $n+1$ 
flavours
\begin{align}
\raisebox{-.5\height}{
  	\begin{tikzpicture}
	\tikzstyle{gauge} = [circle, draw];
	\tikzstyle{flavour} = [regular polygon,regular polygon sides=4, draw];
	\node (g1) [gauge,label=below:{$k$}] {};
	\node (f1) [flavour,above of=g1,label=above:{$n+1$}] {};
	\draw (g1)--(f1) ;
	\end{tikzpicture}
	}
\end{align} 
such that the Higgs branch becomes $\clorbit{(2^k,1^{n+1-2k})}$. To see the 
cotangent bundle of
Grassmann manifolds, start with $k=1$ and recall the $\Ncal=2$ field content 
of SQED: 
the $(n+1)$ $\Ncal=4$ hypermultiplets split into $\Ncal=2$ hypermultiplets 
$X_i$ and $Y_i$, with $i=1,\ldots, n+1$ such that the $\uo$ charges for 
$(X_i,Y_i)$ are $(1,-1)$.
The F and D-term equations are understood as complex and real moment maps 
\begin{align}
 \mu_\C = \sum_{i=1}^{n+1} X_i Y_i \; , \qquad
 \mu_\R = \sum_{i=1}^{n+1} |X_i|^2 -\sum_{i=1}^{n+1} |Y_i|^2
\end{align}
such that the Higgs branch is the hyper-Kähler quotient
\begin{align}
 \MHiggs= (\mu_\R=0,\mu_\C =0) \slash \uo \,.
\end{align}
Turing on a non-trivial real FI parameter $\xi_\R$ leads to $\mu_\R =\xi_\R$, 
which enforces either $X_i=0$ or $Y_i=0$ for all $i=1,\ldots,n+1$, depending on 
the sign of $\xi_\R$. To recognise the $\C P^n$ base, recall that the 
complexified $\uo$ gauge group action identifies $(X_1,\ldots,X_{n+1})\sim 
(\lambda X_1,\ldots, \lambda X_{n+1})$ for $\lambda\in \C^\times = \uo^\C$ and 
$\xi_\R >0$. Moreover, the two different resolutions for $\xi_\R 
\gtreqless 0$ are a manifestation of the basic $A$-type Mukai flop. Hence, 
$(\mu_\R=\xi_\R,\mu_\C =0) \slash \uo \cong T^* 
\C P^n$ describes the resolved Higgs branch as a complex manifold.

Generalising to $k\geq 2$, one repeats the same reasoning for F and D terms and 
observes that the complexified $\urm(k)^\C = \glrm(k,\C)$ gauge transformations 
identify the Higgs branch coordinates to a Grassmann manifold $G_{k,n+1-k}$ or 
$G_{n+1-k,k}$, depending on the sign of the real FI 
parameter. Note that the FI-term measures the size of the Grassmann 
manifold, see also \cite{Forcella:2007wk}.
%
%%%%%%%%%%%%%%%%%%%%%%%%%%%%%%%%%%%%%%%%%%%%%%%%%%%%%%%%%%%%%%%%%%%%%%%%%%%%%%%%
   \section{B-type}
\label{sec:B-type}
The Coulomb branch realisation of $B$-type nilpotent orbit closures via unitary 
quiver gauge theories have been 
constructed in \cite{Hanany:2016gbz}. 
In terms of partitions, there are the following two families for $B_n$ which 
are of height two: firstly, 
$\rho=(2^{2k},1^{2n+1-4k})$ for $4\leq 4k\leq 2n+1$ and, secondly,
$\rho=(3,1^{2n-2})$.
We provide the computational results for Coulomb branches corresponding to 
height two orbits of $B_n$ with $n=2,3,4$ in Table 
\ref{tab:results_B-type_examples}. The remarkable observation is that the HWG 
with fluxes factors neatly into a prefactor times the HWG of the singular case, 
computed in \cite{Hanany:2016gbz}.

\begin{table}[!ht]
\centering
 \begin{tabular}{c|c|c|c}
 \toprule
  $\rho$ & $\dim_\C$ & quiver & HWG with flux \\ \midrule
    $(3)$  & $ 2$ &
 	\raisebox{-.5\height}{
 	\begin{tikzpicture}
	\tikzset{node distance = 0.5cm}
	\tikzstyle{gauge} = [circle, draw,inner sep=2.5pt];
	\tikzstyle{flavour} = [regular polygon,regular polygon sides=4,inner 
sep=2.5pt, draw];
	\node (g1) [gauge,label=below:{$1$}] {};
	\node (f1) [flavour,above of=g1,label=above:{$2$}] {};
	\draw (g1)--(f1);
	\end{tikzpicture}
	}
&$ 
x_1^{p_1+p_2} \left(\mu_1 t \right)^{p_1-p_2} \ \PE \left[\mu_1^2 t^2\right] $
  \\
  \midrule
%  
%%%%%%%%%%%%%%%%%%%%%%%%%%%%%%%%%%%%%%%%%%%%%%%%%%
% 
 $(2^2, 1)$ & $ 4$ &
 \raisebox{-.5\height}{
 \begin{tikzpicture}
	\tikzset{node distance = 0.5cm}
	\tikzstyle{gauge} = [circle, draw,inner sep=2.5pt];
	\tikzstyle{flavour} = [regular polygon,regular polygon sides=4,inner 
sep=2.5pt, draw];
	\node (g1) [gauge,label=below:{$1$}] {};
	\node (g0) [right of =g1,xshift=-0.05cm] {};
	\node (g2) [gauge,right of =g0,xshift=-0.45cm,label=below:{$1$}] {};
	\node (f1) [flavour,above of=g2,label=above:{$1$}] {};
	\draw (g2)--(f1);
	\doublearrow{arrows={}}{(g1) -- (g2)};
	\doublearrow{arrows={-Implies}}{(g1) -- (g0)};	
	\end{tikzpicture}
	}
 & no resolution\\
 $(3, 1^2)$ &  $ 6$ &
 \raisebox{-.5\height}{
 \begin{tikzpicture}
	\tikzset{node distance = 0.5cm}
	\tikzstyle{gauge} = [circle, draw,inner sep=2.5pt];
	\tikzstyle{flavour} = [regular polygon,regular polygon sides=4,inner 
sep=2.5pt, draw];
	\node (g1) [gauge,label=below:{$2$}] {};
	\node (g0) [right of =g1,xshift=-0.05cm] {};
	\node (g2) [gauge,right of =g0,xshift=-0.45cm,label=below:{$1$}] {};
	\node (f1) [flavour,above of=g1,label=above:{$2$}] {};
	\draw (g1)--(f1);
	\doublearrow{arrows={}}{(g1) -- (g2)};
	\doublearrow{arrows={-Implies}}{(g1) -- (g0)};	
	\end{tikzpicture}
 }
 & 
 $
 x_1^{p_1+p_2} \left(\mu_1 t^2\right)^{p_1-p_2} \cdot 
 \PE\left[\mu_2^2 t^2 + \mu_1^2 t^4\right]
 $
 \\
 \midrule
%  
%%%%%%%%%%%%%%%%%%%%%%%%%%%%%%%%%%%%%%%%%%%%%%%%%%%%%%%%%
% 
$(2^2, 1^3)$ &  $ 8$ &
\raisebox{-.5\height}{
\begin{tikzpicture}
	\tikzset{node distance = 0.5cm}
	\tikzstyle{gauge} = [circle, draw,inner sep=2.5pt];
	\tikzstyle{flavour} = [regular polygon,regular polygon sides=4,inner 
sep=2.5pt, draw];
	\node (g1) [gauge,label=below:{$1$}] {};
	\node (g2) [gauge,right of =g1,label=below:{$2$}] {};
	\node (g0) [right of =g2,xshift=-0.05cm] {};
	\node (g3) [gauge,right of =g0,xshift=-0.45cm,label=below:{$1$}] {};
	\node (f1) [flavour,above of=g2,label=above:{$1$}] {};
	\draw (g2)--(f1) (g1)--(g2);
	\doublearrow{arrows={}}{(g2) -- (g3)};
	\doublearrow{arrows={-Implies}}{(g2) -- (g0)};	
	\end{tikzpicture}
    }
& no resolution\\
$(3, 1^4)$ &   $ 10$ &
\raisebox{-.5\height}{
\begin{tikzpicture}
	\tikzset{node distance = 0.5cm}
	\tikzstyle{gauge} = [circle, draw,inner sep=2.5pt];
	\tikzstyle{flavour} = [regular polygon,regular polygon sides=4,inner 
sep=2.5pt, draw];
	\node (g1) [gauge,label=below:{$2$}] {};
	\node (g2) [gauge,right of =g1,label=below:{$2$}] {};
	\node (g0) [right of =g2,xshift=-0.05cm] {};
	\node (g3) [gauge,right of =g0,xshift=-0.45cm,label=below:{$1$}] {};
	\node (f1) [flavour,above of=g1,label=above:{$2$}] {};
	\draw (g1)--(f1) (g1)--(g2);
	\doublearrow{arrows={}}{(g2) -- (g3)};
	\doublearrow{arrows={-Implies}}{(g2) -- (g0)};	
	\end{tikzpicture}
    }
& 
$
x_1^{p_1+p_2} \left(\mu_1 t^2 \right)^{p_1-p_2} \cdot 
\PE \left[\mu_2 t^2 +\mu_1^2 t^4\right]
$
\\
\midrule
% 
%%%%%%%%%%%%%%%%%%%%%%%%%%%%%%%%%%%%%%%%%%%%%%%%% 
% 
$(2^2, 1^5)$ & $ 12$ &
\raisebox{-.5\height}{
\begin{tikzpicture}
	\tikzset{node distance = 0.5cm}
	\tikzstyle{gauge} = [circle, draw,inner sep=2.5pt];
	\tikzstyle{flavour} = [regular polygon,regular polygon sides=4,inner 
sep=2.5pt, draw];
	\node (g1) [gauge,label=below:{$1$}] {};
	\node (g2) [gauge,right of =g1,label=below:{$2$}] {};
	\node (g3) [gauge,right of =g2,label=below:{$2$}] {};
	\node (g0) [right of =g3,xshift=-0.05cm] {};
	\node (g4) [gauge,right of =g0,xshift=-0.45cm,label=below:{$1$}] {};
	\node (f1) [flavour,above of=g2,label=above:{$1$}] {};
	\draw (g2)--(f1) (g1)--(g2) (g2)--(g3);
	\doublearrow{arrows={}}{(g3) -- (g4)};
	\doublearrow{arrows={-Implies}}{(g3) -- (g0)};		
	\end{tikzpicture}
    }
& no resolution\\
$(3, 1^6)$ & $ 14$ &
\raisebox{-.5\height}{
\begin{tikzpicture}
	\tikzset{node distance = 0.5cm}
	\tikzstyle{gauge} = [circle, draw,inner sep=2.5pt];
	\tikzstyle{flavour} = [regular polygon,regular polygon sides=4,inner 
sep=2.5pt, draw];
	\node (g1) [gauge,label=below:{$2$}] {};
	\node (g2) [gauge,right of =g1,label=below:{$2$}] {};
	\node (g3) [gauge,right of =g2,label=below:{$2$}] {};
	\node (g0) [right of =g3,xshift=-0.05cm] {};	
	\node (g4) [gauge,right of =g0,xshift=-0.45cm,label=below:{$1$}] {};
	\node (f1) [flavour,above of=g1,label=above:{$2$}] {};
	\draw (g1)--(f1) (g1)--(g2) (g2)--(g3);
	\doublearrow{arrows={}}{(g3) -- (g4)};
	\doublearrow{arrows={-Implies}}{(g3) -- (g0)};	
	\end{tikzpicture}
}
&
$
 x_1^{p_1+p_2} \left(\mu_1 t^2 \right)^{p_1-p_2} \cdot 
 \PE \left[\mu_2 t^2 + \mu_1^2 t^4 \right]
$
\\
$(2^4, 1)$ &  $ 16$ &
\raisebox{-.5\height}{
\begin{tikzpicture}
	\tikzset{node distance = 0.5cm}
	\tikzstyle{gauge} = [circle, draw,inner sep=2.5pt];
	\tikzstyle{flavour} = [regular polygon,regular polygon sides=4,inner 
sep=2.5pt, draw];
	\node (g1) [gauge,label=below:{$1$}] {};
	\node (g2) [gauge,right of =g1,label=below:{$2$}] {};
	\node (g3) [gauge,right of =g2,label=below:{$3$}] {};
	\node (g0) [right of =g3,xshift=-0.05cm] {};	
	\node (g4) [gauge,right of =g0,xshift=-0.45cm,label=below:{$2$}] {};
	\node (f1) [flavour,above of=g4,label=above:{$1$}] {};
	\draw (g4)--(f1) (g1)--(g2) (g2)--(g3);
	\doublearrow{arrows={}}{(g3) -- (g4)};
	\doublearrow{arrows={-Implies}}{(g3) -- (g0)};	
	\end{tikzpicture}
}
& no resolution\\
  \bottomrule
 \end{tabular}
\caption{Coulomb branch quiver gauge theories for $B$-type algebras: 
realisations of the (closures of the) nilpotent orbits of height 2 for $B_n$ 
with $n=1,2,3,4$. The unphysical $\uo$ in $G_F$ can be eliminated as before: 
introducing an auxiliary fugacity $z_0$ for $\uo \subset G_F$, and imposing 
$z_0^{\sum_j p_j} \prod_{i=1}^n z_i^{r_i}=1$, with $r_i$ the ranks of the 
gauge nodes, leads to $x_1=1$ for $(3,1^{2n-2})$ (and $z_0\equiv 1$). This is 
morally equivalent to setting the sum of fluxes to zero in the HWG.}
\label{tab:results_B-type_examples}
\end{table}

We recall that \cite{Hesselink:1978} provides information on the existence 
and form of the resolution of the nilpotent orbits for $B_n$ with $n=2,3,4$. 
For larger ranks, one can consult \cite[Proposition 3.19]{Fu:2003a}: Let 
$\rho$ be a $B$-type partition of $2n+1$, then there exist a symplectic 
resolution of $\clorbit{\rho}$ (and suitable polarisation) if and only if there 
exist an odd number $q\geq 0$ such that the first $q$ parts of $\rho$ are odd 
and the other parts are even. Applying this criterion to the two height two 
families, we find:  
\begin{compactenum}[(i)]
  \item $\rho=(2^{2k},1^{2n+1-4k})$: There does not exist a resolution, since 
$k>0$.
 \item $\rho=(3, 1^{2n-2})$: There exists a resolution for any $n$, since we 
can choose $q=2n-1$.
 \end{compactenum}
Let us compare this to the monopole formula computations.
\paragraph{Partition $\boldsymbol{\rho=(2^{2k},1^{2n+1-4k})}$.}
For the minimal nilpotent orbit of $B_n$ with partition $(2^2, 1^{2n-3})$, the 
Coulomb branch does not allow for a resolution parameter as there is only a 
single $\uo$ flavour charge. In terms of the monopole formula, a non-trivial 
$\uo$ background flux can be absorbed by a simple redefinition of all charges. 
The non-existence of a resolution for $\clorbit{(2^2, 1^{2n-3})}$ is 
consistent with \cite{Hesselink:1978,Fu:2003a}.

We computed only one other member of the family, namely $(2^4, 1)$, for which 
the monopole formula does not give rise to any resolution parameter. Hence, the 
constructions are consistent with the mathematical results.
\paragraph{Partition $\boldsymbol{\rho=(3, 1^{2n-2})}$.}
The Coulomb branch quivers for the next-to-minimal orbit $\clorbit{(3, 
1^{2n-2})}$ have a $\urm(2)$ flavour node, thus admit for a non-trivial 
resolution parameter. Based on the examples computed, we expect that the 
HWG for the entire family with $n\geq 3$
\begin{align}
 \raisebox{-.5\height}{
\begin{tikzpicture}
	\tikzstyle{gauge} = [circle, draw];
	\tikzstyle{flavour} = [regular polygon,regular polygon sides=4,draw];
	\node (g1) [gauge,label=below:{$2$}] {};
	\node (g2) [gauge,right of =g1,label=below:{$2$}] {};
	\node (g3) [right of =g2,label=below:{}] {$\ldots$};
	\node (g4) [gauge,right of =g3,label=below:{$2$}] {};
	\node (g0) [right of =g4,xshift=-0.3cm] {};	
	\node (g5) [gauge,right of =g0,xshift=-0.7cm,label=below:{$1$}] {};
	\node (f1) [flavour,above of=g1,label=above:{$2$}] {};
	\draw (g1)--(f1) (g1)--(g2) (g2)--(g3) (g3)--(g4);
	\doublearrow{arrows={}}{(g4) -- (g5)};
	\doublearrow{arrows={-Implies}}{(g4) -- (g0)};	
	\end{tikzpicture}
}
\end{align}
is given by
\begin{align}
 \HWG_{(p_1,p_2)}^{(3, 1^{2n-2})} (t^2)=x_1^{p_1+p_2} 
 \left(\mu_1 t^2 \right)^{p_1-p_2} \cdot 
\PE \left[\mu_2 t^2 + \mu_1^2 t^4 \right] \,.
\end{align}
To eliminate the overall $\uo$ factor in $G_F$, one imposes $p_1+p_2=0$, which
reduces the effective flux to $p_1 - p_2 \geq 0$, i.e.\ an $\surm(2)$ 
background charge corresponding to a single resolution parameter.
Again, we can compare the result to the known properties of the resolution
\begin{align}
 \pi_{(3, 1^{2n-2})}:  T^*\left(\tfrac{\sorm(2n+1)}{\sorm(2n-1)\times \sorm(2)} 
\right) \to 
\clorbit{(3, 1^{2n-2})} \; .
\end{align}
The symplectic resolution is given by the cotangent bundle of a Hermitian 
symmetric space. Moreover, the exceptional fibre $\pi_{(3, 
1^{2n-2})}^{-1}(0)\cong 
\tfrac{\sorm(2n+1)}{\sorm(2n-1)\times \sorm(2)}$ is reflected in the prefactor 
$\propto \mu_1$, as the (semi-simple) coadjoint orbit 
$\orbit{\mu_1}^{\mathrm{ss}}$ of the first 
fundamental weight $\mu_1$ of $B_n$ is precisely this Hermitian symmetric 
space. To see this, note that the stabiliser of $\mu_1$ in $\sorm(2n+1)$ is 
$\sorm(2n-1)\times \sorm(2)$.
%%%%%%%%%%%%%%%%%%%%%%%%%%%%%%%%%%%%%%%%%%%%%%%%%%%%%%%%%%%%%%%%%%%%%%%%%%%%%%%%
  \section{C-type}
\label{sec:C-type}
In this section, we investigate the resolutions of nilpotent orbits of $C$-type 
via background charges in the monopole formula for the Coulomb branch quivers. 
The construction via unitary quivers as well as the Hilbert series and HWG have 
been provided in \cite{Hanany:2016gbz}.
Restricting to height two, there is exactly one family of partitions for $C_n$ 
to consider: $\rho= (2^k,1^{2(n-k)})$ for $1\leq k\leq n$. We considered all 
height two cased for $C_n$ with $n=2,3,4$ and summarise the computational 
results in Table 
\ref{tab:results_C-type_examples}.

\begin{table}[!ht]
\centering
 \begin{tabular}{c|c|c|c}
 \toprule
  $\rho$ & $\dim_\C$ & quiver & HWG with flux \\ \midrule
  $(2)$  & $ 2$ &
 	\raisebox{-.5\height}{
 	\begin{tikzpicture}
	\tikzset{node distance = 0.5cm}
	\tikzstyle{gauge} = [circle, draw,inner sep=2.5pt];
	\tikzstyle{flavour} = [regular polygon,regular polygon sides=4,inner 
sep=2.5pt, draw];
	\node (g1) [gauge,label=below:{$1$}] {};
	\node (f1) [flavour,above of=g1,label=above:{$2$}] {};
	\draw (g1)--(f1);
	\end{tikzpicture}
	}
&$ 
x_1^{p_1+p_2} (\mu_1 t)^{p_1-p_2} \ \PE[\mu_1^2 t^2] $
  \\
  \midrule
% 
%%%%%%%%%%%%%%%%%%%%%%%%%%%%%%%%%%% 
% 
$(2, 1^2)$ &  $ 4$ &
\raisebox{-.5\height}{
\begin{tikzpicture}
	\tikzset{node distance = 0.5cm}
	\tikzstyle{gauge} = [circle, draw,inner sep=2.5pt];
	\tikzstyle{flavour} = [regular polygon,regular polygon sides=4,inner 
sep=2.5pt, draw];
	\node (g1) [gauge,label=below:{$1$}] {};
	\node (g0) [right of =g1,xshift=-0.45cm] {};
	\node (g2) [gauge,right of =g0,xshift=-0.05cm,label=below:{$1$}] {};
	\node (f1) [flavour,above of=g1,label=above:{$1$}] {};
	\draw (g1)--(f1);
	\doublearrow{arrows={}}{(g2) -- (g1)};
	\doublearrow{arrows={-Implies}}{(g2) -- (g0)};	
	\end{tikzpicture}
}
& no resolution \\ 
$(2^2)$ & $ 6$ &
\raisebox{-.5\height}{
 	\begin{tikzpicture}
	\tikzset{node distance = 0.5cm}
	\tikzstyle{gauge} = [circle, draw,inner sep=2.5pt];
	\tikzstyle{flavour} = [regular polygon,regular polygon sides=4,inner 
sep=2.5pt, draw];
	\node (g1) [gauge,label=below:{$1$}] {};
	\node (g0) [right of =g1,xshift=-0.45cm] {};
	\node (g2) [gauge,right of =g0,xshift=-0.05cm,label=below:{$2$}] {};
	\node (f1) [flavour,above of=g2,label=above:{$2$}] {};
	\draw (g2)--(f1);
	\doublearrow{arrows={}}{(g2) -- (g1)};
	\doublearrow{arrows={-Implies}}{(g2) -- (g0)};	
	\end{tikzpicture}
}
& 
$x_2^{p_1+p_2} \left(\mu_2 t^2 \right)^{p_1-p_2} \cdot
\PE \left[\mu_1^2 t^2 + \mu_2^2 t^4 \right] $
\\ \midrule
% 
%%%%%%%%%%%%%%%%%%%%%%%%%%%%%%%%%%% 
% 
$(2, 1^4)$ & $ 6$ &
\raisebox{-.5\height}{
\begin{tikzpicture}
	\tikzset{node distance = 0.5cm}
	\tikzstyle{gauge} = [circle, draw,inner sep=2.5pt];
	\tikzstyle{flavour} = [regular polygon,regular polygon sides=4,inner 
sep=2.5pt, draw];
	\node (g1) [gauge,label=below:{$1$}] {};
	\node (g2) [gauge,right of =g1,label=below:{$1$}] {};
	\node (g0) [right of =g2,xshift=-0.45cm] {};	
	\node (g3) [gauge,right of =g0,xshift=-0.05cm,label=below:{$1$}] {};
	\node (f1) [flavour,above of=g1,label=above:{$1$}] {};
	\draw (g1)--(f1) (g1)--(g2);
	\doublearrow{arrows={}}{(g3) -- (g2)};
	\doublearrow{arrows={-Implies}}{(g3) -- (g0)};	
	\end{tikzpicture}
}
& no resolution \\ 
$(2^2, 1^2)$ &  $ 10$ &
\raisebox{-.5\height}{
\begin{tikzpicture}
	\tikzset{node distance = 0.5cm}
	\tikzstyle{gauge} = [circle, draw,inner sep=2.5pt];
	\tikzstyle{flavour} = [regular polygon,regular polygon sides=4,inner 
sep=2.5pt, draw];
	\node (g1) [gauge,label=below:{$1$}] {};
	\node (g2) [gauge,right of =g1,label=below:{$2$}] {};
	\node (g0) [right of =g2,xshift=-0.45cm] {};	
	\node (g3) [gauge,right of =g0,xshift=-0.05cm,label=below:{$2$}] {};
	\node (f1) [flavour,above of=g2,label=above:{$1$}] {};
	\draw (g2)--(f1) (g1)--(g2);
	\doublearrow{arrows={}}{(g3) -- (g2)};
	\doublearrow{arrows={-Implies}}{(g3) -- (g0)};	
	\end{tikzpicture}
}
& no resolution \\
$(2^3)$ & $ 12$ &
\raisebox{-.5\height}{
\begin{tikzpicture}
	\tikzset{node distance = 0.5cm}
	\tikzstyle{gauge} = [circle, draw,inner sep=2.5pt];
	\tikzstyle{flavour} = [regular polygon,regular polygon sides=4,inner 
sep=2.5pt, draw];
	\node (g1) [gauge,label=below:{$1$}] {};
	\node (g2) [gauge,right of =g1,label=below:{$2$}] {};
	\node (g0) [right of =g2,xshift=-0.45cm] {};	
	\node (g3) [gauge,right of =g0,xshift=-0.05cm,label=below:{$3$}] {};
	\node (f1) [flavour,above of=g3,label=above:{$2$}] {};
	\draw (g3)--(f1) (g1)--(g2);
	\doublearrow{arrows={}}{(g3) -- (g2)};
	\doublearrow{arrows={-Implies}}{(g3) -- (g0)};	
	\end{tikzpicture}
}
& 
$  x_3^{p_1 +p_2} \left(\mu_3 t^{3} \right)^{p_1 -p_2} \cdot 
\PE \left[\mu_1^2 t^2 + \mu_2^2 t^4 + \mu_3^2 t^6 \right] $
\\
 \midrule
% 
%%%%%%%%%%%%%%%%%%%%%%%%%%%%%%%%%%%%%%%%%%%%5 
% 
$(2, 1^6)$ & $ 8$ &
\raisebox{-.5\height}{
\begin{tikzpicture}
	\tikzset{node distance = 0.5cm}
	\tikzstyle{gauge} = [circle, draw,inner sep=2.5pt];
	\tikzstyle{flavour} = [regular polygon,regular polygon sides=4,inner 
sep=2.5pt, draw];
	\node (g1) [gauge,label=below:{$1$}] {};
	\node (g2) [gauge,right of =g1,label=below:{$1$}] {};
	\node (g3) [gauge,right of =g2,label=below:{$1$}] {};
	\node (g0) [right of =g3,xshift=-0.45cm] {};	
	\node (g4) [gauge,right of =g0,xshift=-0.05cm,label=below:{$1$}] {};
	\node (f1) [flavour,above of=g1,label=above:{$1$}] {};
	\draw (g1)--(f1) (g1)--(g2) (g2)--(g3);
	\doublearrow{arrows={}}{(g4) -- (g3)};
	\doublearrow{arrows={-Implies}}{(g4) -- (g0)};		
	\end{tikzpicture}
}
& no resolution \\ 
$(2^2, 1^4)$ & $ 14$ &
\raisebox{-.5\height}{
\begin{tikzpicture}
	\tikzset{node distance = 0.5cm}
	\tikzstyle{gauge} = [circle, draw,inner sep=2.5pt];
	\tikzstyle{flavour} = [regular polygon,regular polygon sides=4,inner 
sep=2.5pt, draw];
	\node (g1) [gauge,label=below:{$1$}] {};
	\node (g2) [gauge,right of =g1,label=below:{$2$}] {};
	\node (g3) [gauge,right of =g2,label=below:{$2$}] {};
	\node (g0) [right of =g3,xshift=-0.45cm] {};	
	\node (g4) [gauge,right of =g0,xshift=-0.05cm,label=below:{$2$}] {};
	\node (f1) [flavour,above of=g2,label=above:{$1$}] {};
	\draw (g2)--(f1) (g1)--(g2) (g2)--(g3);
	\doublearrow{arrows={}}{(g4) -- (g3)};
	\doublearrow{arrows={-Implies}}{(g4) -- (g0)};	
	\end{tikzpicture}
}
& no resolution \\
$(2^3, 1^2)$ & $ 18$ &
\raisebox{-.5\height}{
\begin{tikzpicture}
	\tikzset{node distance = 0.5cm}
	\tikzstyle{gauge} = [circle, draw,inner sep=2.5pt];
	\tikzstyle{flavour} = [regular polygon,regular polygon sides=4,inner 
sep=2.5pt, draw];
	\node (g1) [gauge,label=below:{$1$}] {};
	\node (g2) [gauge,right of =g1,label=below:{$2$}] {};
	\node (g3) [gauge,right of =g2,label=below:{$3$}] {};
	\node (g0) [right of =g3,xshift=-0.45cm] {};	
	\node (g4) [gauge,right of =g0,xshift=-0.05cm,label=below:{$3$}] {};
	\node (f1) [flavour,above of=g3,label=above:{$1$}] {};
	\draw (g3)--(f1) (g1)--(g2) (g2)--(g3);
	\doublearrow{arrows={}}{(g4) -- (g3)};
	\doublearrow{arrows={-Implies}}{(g4) -- (g0)};	
	\end{tikzpicture}
}
& no resolution \\
$(2^4)$ & $ 20$ &
\raisebox{-.5\height}{
\begin{tikzpicture}
	\tikzset{node distance = 0.5cm}
	\tikzstyle{gauge} = [circle, draw,inner sep=2.5pt];
	\tikzstyle{flavour} = [regular polygon,regular polygon sides=4,inner 
sep=2.5pt, draw];
	\node (g1) [gauge,label=below:{$1$}] {};
	\node (g2) [gauge,right of =g1,label=below:{$2$}] {};
	\node (g3) [gauge,right of =g2,label=below:{$3$}] {};
	\node (g0) [right of =g3,xshift=-0.45cm] {};	
	\node (g4) [gauge,right of =g0,xshift=-0.05cm,label=below:{$4$}] {};
	\node (f1) [flavour,above of=g4,label=above:{$2$}] {};
	\draw (g4)--(f1) (g1)--(g2) (g2)--(g3);
	\doublearrow{arrows={}}{(g4) -- (g3)};
	\doublearrow{arrows={-Implies}}{(g4) -- (g0)};	
	\end{tikzpicture}
}
& 
$ x_4^{p_1+p_2} \left(\mu_4 t^4 \right)^{p_1-p_2} \cdot 
\PE \left[\mu_1^2 t^2 + \mu_2^2 t^4 + \mu_3^2 t^6 + \mu_4^2 t^8 \right] 
$
\\
  \bottomrule
 \end{tabular}
\caption{Coulomb branch quiver gauge theories for $C$-type algebras: 
realisations of the (closures of the) nilpotent orbits of height 2 for $C_n$, 
for $n=1,2,3,4$. To eliminate the unphysical $\uo$ in $G_F$ in the HWG, one 
proceeds as before: 
introducing an auxiliary fugacity $z_0$ for $\uo \subset G_F$, and imposing 
$z_0^{\sum_j p_j} \prod_{i=1}^n z_i^{r_i}=1$, with $r_i$ the ranks of the 
gauge nodes, leads to $x_n=1$ for $(2^{n})$ (and $z_0\equiv 1$). This is 
effectively the same as setting the sum of fluxes to zero in the HWG.}
\label{tab:results_C-type_examples}
\end{table}

As in the previous case, existence and form of the resolution have been 
tabulated in \cite{Hesselink:1978} for low rank cases. The general criterion 
has been formulated in \cite[Proposition 3.19]{Fu:2003a}:  Let $\rho$ be a 
$C$-type 
partition of $2n$, then there exist a symplectic 
resolution of $\clorbit{\rho}$ (and suitable polarisation) if and only if there 
exists an even number $q\geq 0$ such that the first $q$ parts of $\rho$ are odd 
and the other parts are even. Inspecting the height two family 
$\rho=(2^k,1^{2(n-k)})$ we find:
there exists a resolution only for $n=k$, as we then choose $q=0$. Hence, 
$\rho=(2^n)$ admits a resolution and $\rho=(2^k,1^{2(n-k)})$ with $n>k\geq 1$ 
does not. 
\paragraph{Partition $\boldsymbol{\rho=(2^k,1^{2(n-k)})}$, $\boldsymbol{n>k\geq 
1}$.}
For the minimal nilpotent orbit $\clorbit{(2, 1^{2n-2})}$ of $C_n$, the Coulomb 
branch quivers do not allow for a resolution parameter, as there is only a 
single $\uo$ flavour node present. 
Next, the orbits of partition $(2^2, 1^{2n-4})$, $n\geq 3$ do not admit a 
resolution. The Coulomb branch quiver construction are consistent with this.
In addition, the quivers corresponding to the partition $(2^3, 1^{2n-6})$ 
do not give rise to any resolution, because the only available flavour is 
a $\uo$ node. 
Hence, the considered examples do agree with 
\cite{Hesselink:1978,Fu:2003a}.
\paragraph{Partition $\boldsymbol{\rho=(2^n)}$.}
Lastly, the orbits of partition $(2^n)$ for $C_n$ exhibit a $\urm(2)$ flavour; 
thus, a non-trivial resolution parameter exists. From the examples computed, we 
expect that the HWG for the entire family
\begin{align}
 \raisebox{-.5\height}{
\begin{tikzpicture}
	\tikzstyle{gauge} = [circle, draw];
	\tikzstyle{flavour} = [regular polygon,regular polygon sides=4,draw];
	\node (g1) [gauge,label=below:{$1$}] {};
	\node (g2) [gauge,right of =g1,label=below:{$2$}] {};
	\node (g3) [right of =g2,label=below:{}] {$\ldots$};
	\node (g4) [gauge,right of =g3,label=below:{$n{-}1$}] {};
	\node (g0) [right of =g4,xshift=-0.7cm] {};	
	\node (g5) [gauge,right of =g0,xshift=-0.3cm,label=below:{$n$}] {};
	\node (f1) [flavour,above of=g5,label=above:{$2$}] {};
	\draw (g5)--(f1) (g1)--(g2) (g2)--(g3) (g3)--(g4);
	\doublearrow{arrows={}}{(g5) -- (g4)};
	\doublearrow{arrows={-Implies}}{(g5) -- (g0)};	
	\end{tikzpicture}
}
\end{align}
is given by
\begin{align}
 \HWG_{(p_1,p_2)}^{(2^n)}(t^2) =x_n^{p_1+p_2}  \left(\mu_n t^{n} 
\right)^{p_1-p_2} 
\cdot \PE\left[\sum_{i=1}^n \mu_i^2 t^{2i} \right] \; .
\end{align}
In order to eliminate the overall $\uo$ in $G_F$, one imposes $p_1+p_2=0$.
This reduces the fluxes to a single  $p_1 - p_2\geq 0$ background charge of 
$\surm(2)$, which corresponds to exactly one resolution parameter. Comparing the 
result to the known resolution
\begin{align}
 \pi_{(2^n)}: T^* \left( \tfrac{\sprm(n)}{\urm(n)} \right) \to \clorbit{(2^n)} 
\; ,
\end{align}
we observe again the cotangent bundle of a Hermitian symmetric space, see 
Table \ref{tab:HSS}. In 
addition, the exceptional fibre $\pi_{(2^n)}^{-1}(0) \cong 
\tfrac{\sprm(n)}{\urm(n)} 
$ is reflected in the HWG by the prefactor $\propto \mu_n$. To see this, recall 
that the stabiliser of the $n$-th fundamental weight $\mu_n$ of $C_n$ is 
given by $\urm(n)\cong \surm(n)\times \uo$. Hence, the (semi-simple) coadjoint 
orbit 
$\orbit{\mu_n}^{\mathrm{ss}}$ through $\lambda_n$ is isomorphic to the 
Hermitian symmetric space $\tfrac{\sprm(n)}{ \urm(n)}$.
\paragraph{Remarks.}
As a consistency check, one observes that the accidental isomorphism $B_2 \cong 
C_2$ is respected by the Coulomb branch computations. For instance, the 
$B$-type $(3,1^2)$ of Table \ref{tab:results_B-type_examples} agrees with 
$C$-type $(2^2)$ of Table \ref{tab:results_C-type_examples} upon identification 
of fugacity and weight labels. Similarly, the isomorphism $A_1 \cong B_1 \cong 
C_1$ is manifest in the results.
% 
%%%%%%%%%%%%%%%%%%%%%%%%%%%%%%%%%%%%%%%%%%%%%%%%%%%%%%%%%%%%%%%%%%%%%%%%%%%%%%%%
  \section{D-type}
\label{sec:D-type}
The last classical case to consider is the nilpotent orbits of $D$-type. The 
Coulomb branch construction of the unitary quiver as well as the Hilbert series 
and HWG has been given in \cite{Hanany:2016gbz}. Here, we analyse the 
resolutions via background charges in the monopole formula.
Restricting ourselves to height two, there are only two families of partitions 
for $D_n$ to consider: firstly, $\rho= (2^{2k},1^{2n-4k})$ for $2\leq 2k \leq 
n$ and, secondly, $\rho=(3,1^{2n-3})$ for $n\geq 2$.
The details of the height two cases considered for $D_n$ with $n=3,4,5$ are 
provided in Table
\ref{tab:results_D-type_examples}.

\begin{table}[!ht]
\centering
 \begin{tabular}{c|c|c|c}
 \toprule
  $\rho$ & $\dim_\C$  & quiver & HWG with flux \\ \midrule
  $(2^2, 1^2 )$ & $ 6$ &
  \raisebox{-.5\height}{
  \begin{tikzpicture}
	\tikzset{node distance = 0.5cm}
	\tikzstyle{gauge} = [circle, draw,inner sep=2.5pt];
	\tikzstyle{flavour} = [regular polygon,regular polygon sides=4,inner 
sep=2.5pt, draw];
	\node (g1) [gauge,label=below:{$1$}] {};
	\node (g6) [gauge,below right of =g1,label=below:{$1$}] {};
	\node (g7) [gauge,above right of =g1,label=above:{$1$}] {};
	\node (f1) [flavour,right of=g6,label=below:{$1$}] {};
	\node (f2) [flavour,right of=g7,label=above:{$1$}] {};
	\draw  (g1)--(g6) (f1)--(g6) 
(g1)--(g7) (f2)--(g7);
	\end{tikzpicture}
  }
  & 
  $
  \begin{cases}
   x_2^{p_2} x_3^{p_3} \left(\mu_3 t \right)^{p_2 -p_3} \cdot
   \PE\left[\mu_2 \mu_3 t^2\right] & p_2 \geq p_3 \\
   x_2^{p_2} x_3^{p_3} \left(\mu_2 t \right)^{p_3 -p_2} \cdot
   \PE \left[\mu_2 \mu_3 t^2 \right] & p_2 \leq p_3
   \end{cases}
  $  
  \\
  $(3,  1^3 )$ & $ 8$ &
   \raisebox{-.5\height}{
  \begin{tikzpicture}
	\tikzset{node distance = 0.5cm}
	\tikzstyle{gauge} = [circle, draw,inner sep=2.5pt];
	\tikzstyle{flavour} = [regular polygon,regular polygon sides=4,inner 
sep=2.5pt, draw];
	\node (g1) [gauge,label=below:{$2$}] {};
	\node (g6) [gauge,below right of =g1,label=right:{$1$}] {};
	\node (g7) [gauge,above right of =g1,label=right:{$1$}] {};
	\node (f1) [flavour,left of=g1,label=below:{$2$}] {};
	\draw  (g1)--(g6) (g1)--(g7) (g1)--(f1);
	\end{tikzpicture}
  }
  & 
  $ x_1^{p_1+p_2} \left(\mu_1 t^2 \right)^{p_1 -p_2} \cdot 
  \PE\left[\mu_1^2 t^4 +\mu_2 \mu_3 t^2\right]
  $
  \\ \midrule
%   
%%%%%%%%%%%%%%%%%%%%%%%%%%%%%%%%%%%
% 
  $(2^2, 1^4 )$ &  $ 10$ &
  \raisebox{-.5\height}{
  	\begin{tikzpicture}
	\tikzset{node distance = 0.5cm}
	\tikzstyle{gauge} = [circle, draw,inner sep=2.5pt];
	\tikzstyle{flavour} = [regular polygon,regular polygon sides=4,inner 
sep=2.5pt, draw];
	\node (g1) [gauge,label=left:{$1$}] {};
	\node (g2) [gauge, below right of =g1,label=above:{$2$}] {};
	\node (g3) [gauge,above right of =g2,label=right:{$1$}] {};
	\node (g4) [gauge,below right of =g2,label=right:{$1$}] {};
	\node (f1) [flavour,below left of=g2,label=left:{$1$}] {};
	\draw (g1)--(g2) (g3)--(g2) (g4)--(g2) (f1)--(g2);
	\end{tikzpicture}
  }
  & no resolution\\
  $(3,  1^5 )$ &  $ 12$ &
  \raisebox{-.5\height}{
  \begin{tikzpicture}
	\tikzset{node distance = 0.5cm}
	\tikzstyle{gauge} = [circle, draw,inner sep=2.5pt];
	\tikzstyle{flavour} = [regular polygon,regular polygon sides=4,inner 
sep=2.5pt, draw];
	\node (g1) [gauge,label=below:{$2$}] {};
	\node (g2) [gauge, right of =g1,label=below:{$2$}] {};
	\node (g3) [gauge,above right of =g2,label=right:{$1$}] {};
	\node (g4) [gauge,below right of =g2,label=right:{$1$}] {};
	\node (f1) [flavour,left of=g1,label=below:{$2$}] {};
	\draw (g1)--(g2) (g3)--(g2) (g4)--(g2) (f1)--(g1);
	\end{tikzpicture}
  }
  & 
  $  x_1^{p_1+p_2} \left(\mu_1 t^2 \right)^{p_1 -p_2} \cdot 
  \PE \left[\mu_1^2 t^4 +\mu_2 t^2 \right]   $
  \\
    $(2^4)^I$ &  $ 12$ &
  \raisebox{-.5\height}{
  \begin{tikzpicture}
	\tikzset{node distance = 0.5cm}
	\tikzstyle{gauge} = [circle, draw,inner sep=2.5pt];
	\tikzstyle{flavour} = [regular polygon,regular polygon sides=4,inner 
sep=2.5pt, draw];
	\node (g1) [gauge,label=below:{$1$}] {};
	\node (g2) [gauge, right of =g1,label=below:{$2$}] {};
	\node (g3) [gauge,below right of =g2,label=right:{$1$}] {};
	\node (g4) [gauge,above right of =g2,label=above:{$2$}] {};
	\node (f1) [flavour,right of=g4,label=above:{$2$}] {};
	\draw (g1)--(g2) (g3)--(g2) (g4)--(g2) (f1)--(g4);
	\end{tikzpicture}
  }
  & $x_3^{p_1+p_2} \left(\mu_3 t^2 \right)^{p_1 -p_2} \cdot 
  \PE \left[\mu_3^2 t^4 +\mu_2 t^2 \right]$  \\
  $(2^4)^{II}$ &  $ 12$ &
  \raisebox{-.5\height}{
  \begin{tikzpicture}
	\tikzset{node distance = 0.5cm}
	\tikzstyle{gauge} = [circle, draw,inner sep=2.5pt];
	\tikzstyle{flavour} = [regular polygon,regular polygon sides=4,inner 
sep=2.5pt, draw];
	\node (g1) [gauge,label=below:{$1$}] {};
	\node (g2) [gauge, right of =g1,label=below:{$2$}] {};
	\node (g3) [gauge,above right of =g2,label=right:{$1$}] {};
	\node (g4) [gauge,below right of =g2,label=below:{$2$}] {};
	\node (f1) [flavour,right of=g4,label=below:{$2$}] {};
	\draw (g1)--(g2) (g3)--(g2) (g4)--(g2) (f1)--(g4);
	\end{tikzpicture}
  }
  & $x_4^{p_1+p_2} \left(\mu_4 t^2 \right)^{p_1 -p_2} \cdot 
  \PE \left[\mu_4^2 t^4 +\mu_2 t^2 \right]$ \\ 
\midrule
%   
%%%%%%%%%%%%%%%%%%%%%%%%%%%%%%%%%%%
% 
  $(2^2, 1^6 )$ &  $14$ &
  \raisebox{-.5\height}{
  \begin{tikzpicture}
	\tikzset{node distance = 0.5cm}
	\tikzstyle{gauge} = [circle, draw,inner sep=2.5pt];
	\tikzstyle{flavour} = [regular polygon,regular polygon sides=4,inner 
sep=2.5pt, draw];
	\node (g1) [gauge,label=below:{$1$}] {};
	\node (g2) [gauge, right of =g1,label=below:{$2$}] {};
	\node (g3) [gauge,right of =g2,label=below:{$2$}] {};
	\node (g4) [gauge,below right of =g3,label=right:{$1$}] {};
	\node (g5) [gauge,above right of =g3,label=right:{$1$}] {};
	\node (f1) [flavour,above of=g2,label=above:{$1$}] {};
	\draw  (g1)--(g2) (g2)--(g3) (g3)--(g4) (f1)--(g2) 
(g3)--(g5);
	\end{tikzpicture}
  }
  & no resolution \\
  $(3,  1^7 )$ &  $ 16$ &
  \raisebox{-.5\height}{
  \begin{tikzpicture}
	\tikzset{node distance = 0.5cm}
	\tikzstyle{gauge} = [circle, draw,inner sep=2.5pt];
	\tikzstyle{flavour} = [regular polygon,regular polygon sides=4,inner 
sep=2.5pt, draw];
	\node (g1) [gauge,label=below:{$2$}] {};
	\node (g2) [gauge, right of =g1,label=below:{$2$}] {};
	\node (g3) [gauge, right of =g2,label=below:{$2$}] {};
	\node (g4) [gauge,above right of =g3,label=right:{$1$}] {};
	\node (g5) [gauge,below right of =g3,label=right:{$1$}] {};
	\node (f1) [flavour,left of=g1,label=below:{$2$}] {};
	\draw (g1)--(g2) (g3)--(g2) (g4)--(g3) (g5)--(g3) (f1)--(g1);
	\end{tikzpicture}
  }
  &$x_1^{p_1+p_2} \left(\mu_1 t^2 \right)^{p_1 -p_2} \cdot 
  \PE \left[\mu_1^2 t^4 + \mu_2 t^2 \right]$\\ 
  $(2^4,  1^2 )$ & $ 20$ &
  \raisebox{-.5\height}{
  \begin{tikzpicture}
	\tikzset{node distance = 0.5cm}
	\tikzstyle{gauge} = [circle, draw,inner sep=2.5pt];
	\tikzstyle{flavour} = [regular polygon,regular polygon sides=4,inner 
sep=2.5pt, draw];
	\node (g1) [gauge,label=below:{$1$}] {};
	\node (g4) [gauge, right of =g1,label=below:{$2$}] {};
	\node (g5) [gauge,right of =g4,label=below:{$3$}] {};
	\node (g6) [gauge,below right of =g5,label=below:{$2$}] {};
	\node (g7) [gauge,above right of =g5,label=above:{$2$}] {};
	\node (f1) [flavour,right of=g6,label=below:{$1$}] {};
	\node (f2) [flavour,right of=g7,label=above:{$1$}] {};
	\draw  (g1)--(g4) (g4)--(g5) (g5)--(g6) (f1)--(g6) 
(g5)--(g7) (f2)--(g7);
	\end{tikzpicture}
  }
  & 
  $\begin{cases}
  x_4^{p_4} x_5^{p_5} \left(\mu_5 t^2 \right)^{p_4 -p_5} \cdot 
  \PE \left[\mu_2^2 t^2+ \mu_4 \mu_5 t^4 \right] & p_4 \geq p_5 \\
  x_4^{p_4} x_5^{p_5} \left(\mu_4 t^2 \right)^{p_5 -p_4} \cdot  
  \PE \left[\mu_2^2 t^2+ \mu_4 \mu_5 t^4 \right] & p_4 \leq p_5
  \end{cases}
  $
  \\
    \bottomrule
 \end{tabular}
\caption{Coulomb branch quiver gauge theories for $D$-type algebras: 
realisations of the (closures of the) nilpotent orbits of height 2 for $D_n$ 
with $n=3,4,5$. To eliminate the unphysical $\uo$ in $G_F$ one repeats the 
earlier arguments. 
Introducing an auxiliary fugacity $z_0$ for $\uo \subset G_F$, and imposing 
$z_0^{\sum_j p_j} \prod_{i=1}^n z_i^{r_i}=1$, with $r_i$ the ranks of the 
gauge nodes, leads to the following cases:
$x_1=1$ for $(3,1^{2n-3})$, 
$x_{n-1} x_{n}=1$ for $(2^{n-1},1^2)$ and $n=$ odd, and  $x_{n-1}=1 $ or $x_n=1 
$ for $(2^n)$ and $n=$ even (and $z_0\equiv 1$).}
\label{tab:results_D-type_examples}
\end{table}

The existence and form of polarisations for low rank $D$-type nilpotent orbits 
is tabulated in \cite{Hesselink:1978}. For the general statement, we refer to 
\cite[Proposition 3.20]{Fu:2003a}: Let $\rho$ be a $D$-type 
partition of $2n$, then there exist a symplectic 
resolution of $\clorbit{\rho}$ (and suitable polarisation) if and only if 
either there exists an even number $q\neq 2$ such that the first $q$ parts of 
$\rho$ are odd and the other parts are even, or there exist exactly $2$ odd 
parts which are at position $2k-1$ and $2k$ in $\rho$ for some $k$. Inspecting 
the two height two families, we find the following:
\begin{compactenum}[(i)]
 \item $\rho=(2^{2k},1^{2n-4k})$: There exists a resolution either for 
$2k=n-1$, i.e.\ for $(2^{n-1},1^2)$ of $D_n$ with $n=$ odd, or for 
$2k=n$, i.e.\ for $(2^{n})$ of $D_n$ with $n=$ even, which is a very even 
partition. For all other choices of $k$ there does not exist a symplectic 
resolution.
  \item $\rho=(3,1^{2n-3})$: There exists a resolution for any $n$.
\end{compactenum}
Let us compare this to the monopole formula computations.
\paragraph{Partition $\boldsymbol{\rho=(2^{2k},1^{2n-4k})}$, 
$\boldsymbol{n-1>2k\geq2}$.}
To begin with, consider the minimal nilpotent orbit $\orbit{(2^2, 1^{2n-4})}$ 
of 
$D_n$. The Coulomb branch quiver 
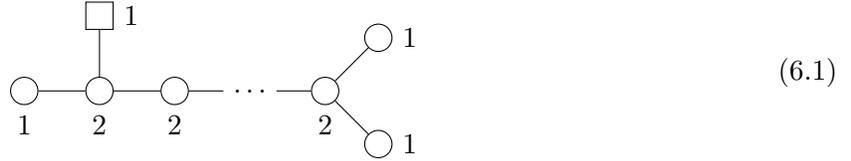
\begin{align}
\raisebox{-.5\height}{
 	\begin{tikzpicture}
	\tikzstyle{gauge} = [circle, draw];
	\tikzstyle{flavour} = [regular polygon,regular polygon sides=4, draw];
	\node (g1) [gauge,label=below:{$1$}] {};
	\node (g2) [gauge, right of =g1,label=below:{$2$}] {};
	\node (g3) [gauge,right of =g2,label=below:{$2$}] {};
	\node (gg) [right of =g3] {$\ldots$};
	\node (g4) [gauge, right of =gg,label=below:{$2$}] {};
	\node (g5) [gauge,above right of =g4,label=right:{$1$}] {};
	\node (g6) [gauge,below right of =g4,label=right:{$1$}] {};
	\node (f1) [flavour,above of=g2,label=right:{$1$}] {};
	\draw (g1)--(g2) (g3)--(g2) (g3)--(gg) (g4)--(gg) (g4)--(g5) (g4)--(g6) 
(f1)--(g2);
	\end{tikzpicture}
	}
\end{align}
does not allow for any non-trivial resolution parameter. 
The monopole formula is consistent with the results of 
\cite{Hesselink:1978,Fu:2003a}.
\paragraph{Partition $\boldsymbol{\rho=(3,1^{2n-3})}$.}
Next, we consider $\clorbit{(3, 1^{2n-3})}$ of $D_n$, $n\geq 4$, for which the 
Coulomb branch quiver is the following: 
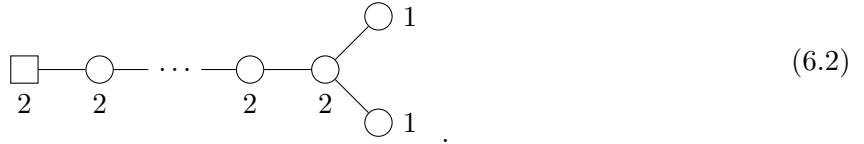
\begin{align}
\raisebox{-.5\height}{
 	\begin{tikzpicture}
	\tikzstyle{gauge} = [circle, draw];
	\tikzstyle{flavour} = [regular polygon,regular polygon sides=4, draw];
	\node (g1) [gauge,label=below:{$2$}] {};
	\node (g2) [right of=g1]{$\ldots$};
	\node (g3) [gauge, right of =g2,label=below:{$2$}] {};
	\node (g4) [gauge, right of =g3,label=below:{$2$}] {};
	\node (g5) [gauge,above right of =g4,label=right:{$1$}] {};
	\node (g6) [gauge,below right of =g4,label=right:{$1$}] {};
	\node (f1) [flavour,left of=g1,label=below:{$2$}] {};
	\draw (g1)--(g2) (g3)--(g2) (g4)--(g3) (g5)--(g4) (g6)--(g4) (f1)--(g1);
	\end{tikzpicture} \,.
	}
\end{align}
Based on the examples computed, we expect that the HWG for the entire family is 
given by
\begin{align}
  \HWG_{(p_1,p_2)}^{(3,1^{2n-3})} (t^2) =x_1^{p_1 +p_2}   
  \left(\mu_1 t^2 \right)^{p_1 -p_2} 
\cdot \PE\left[\mu_2 t^2 + \mu_1^2 t^4 \right] \,.
\end{align}
One can eliminate the overall shift symmetry in the fluxes via the condition 
$p_1+p_2=0$ such that one obtains a single $\surm(2)$ background charge $p_1-p_2 
\geq0$, which corresponds to one effective resolution parameter.
The structure of the results suggest to compare it to the known resolution
\begin{align}
 \pi_{(3, 1^{2n-3})} : T^* \left(\tfrac{\sorm(2n)}{\sorm(2n-2)\times \sorm(2)} 
\right) \to 
\clorbit{(3, 1^{2n-3})}
\end{align}
for which the exceptional fibre is 
$\pi_{(3, 1^{2n-3})}^{-1}(0)\cong \tfrac{\sorm(2n)}{\sorm(2n-2)\times 
\sorm(2)}$. The prefactor 
$\propto \mu_1$ indicates this, because the stabiliser of the first fundamental 
weight $\mu_1$ of $D_n$ is $\sorm(2n-2)\times \uo$. Hence, the (semi-simple) 
coadjoint orbit $\orbit{\mu_1}^{\mathrm{ss}}$ through $\mu_1$ is isomorphic 
to the HSS $\tfrac{\sorm(2n)}{\sorm(2n-2)\times \sorm(2)}$.
\paragraph{Partition $\boldsymbol{\rho=(2^{n-1},1^2)}$, $\boldsymbol{n=}$ odd.}
Consider the orbit closure $\clorbit{(2^{n-1}, 1^2)}$ of $D_n\equiv 
D_{2l+1}$ via its 
Coulomb branch realisation
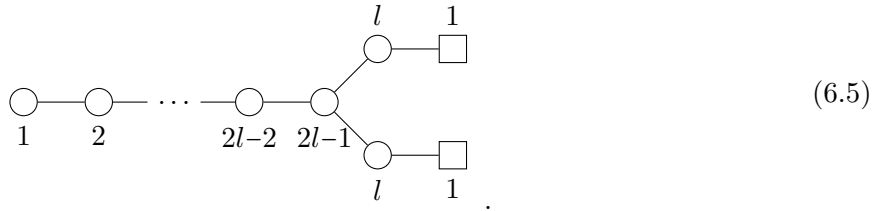
\begin{align}
\raisebox{-.5\height}{
 	\begin{tikzpicture}
	\tikzstyle{gauge} = [circle, draw];
	\tikzstyle{flavour} = [regular polygon,regular polygon sides=4, draw];
	\node (g1) [gauge,label=below:{$1$}] {};
	\node (g2) [gauge, right of =g1,label=below:{$2$}] {};
	\node (g3) [ right of =g2] {$\ldots$};
	\node (g4) [gauge, right of =g3,label=below:{$2l{-}2$}] {};
	\node (g5) [gauge,right of =g4,label=below:{$2l{-}1$}] {};
	\node (g6) [gauge,below right of =g5,label=below:{$l$}] {};
	\node (g7) [gauge,above right of =g5,label=above:{$l$}] {};
	\node (f1) [flavour,right of=g6,label=below:{$1$}] {};
	\node (f2) [flavour,right of=g7,label=above:{$1$}] {};
	\draw (g1)--(g2) (g2)--(g3) (g3)--(g4) (g4)--(g5) (g5)--(g6) (f1)--(g6) 
(g5)--(g7) (f2)--(g7);
	\end{tikzpicture} \,.
	}
\end{align}
As apparent from the quiver, the Coulomb branch allows for a resolution 
parameter and based on the examples computed, we anticipate the HWG to be
\begin{align}
 \HWG_{(p_{2l},p_{2l+1})}^{(2^{n-1},1^2)} (t^2) = \begin{cases}
     x_{2l}^{p_{2l}} x_{2l+1}^{p_{2l+1}}
     \left(\mu_{2l+1} t^2 \right)^{p_{2l}-p_{2l+1}} 
     \cdot \PE\left[\sum_{i=1}^{l-1} \mu_{2i}t^{2i} +\mu_{2l} 
\mu_{2l+1}t^{n-1}\right]
     & , p_{2l}\geq p_{2l+1} \\
     x_{2l}^{p_{2l}} x_{2l+1}^{p_{2l+1}}
     \left(\mu_{2l} t^2 \right)^{p_{2l}-p_{2l+1}} 
     \cdot \PE\left[\sum_{i=1}^{l-1} \mu_{2i}t^{2i} +\mu_{2l} 
\mu_{2l+1}t^{n-1}\right]
     &, p_{2l} \leq p_{2l+1} 
        \end{cases}\,.
\end{align}
As before, one can eliminate the overall $\uo$ in $G_F$ by imposing 
$p_{2l}+p_{2l+1}=0$, which reduces the system to a single (positive) resolution 
parameter $\sim \pm(p_{2l}-p_{2l+1})$.
Comparing this to the literature \cite{Hesselink:1978}, the resolution is of 
the form
\begin{align}
 \pi_{(2^{n-1}, 1^2)} : T^* \left( \tfrac{\sorm(2n)}{\surm(n)\times\uo} \right) 
\to 
\clorbit{(2^{n-1}, 1^2)} \;,
\end{align}
with exceptional fibre $\pi_{(2^{n-1}, 1^2)}^{-1}(0) \cong 
\tfrac{\sorm(2n)}{\surm(n)\times\uo}$. As apparent from the HWG, there are 
again two cases for the resolution, depending on the relative sign of 
$p_{2l}- p_{2l+1}$. This is a manifestation of the Mukai flop of $D$-type, see 
\eqref{eq:flop_D-type}.
\paragraph{Partition $\boldsymbol{\rho=(2^{n})}$, $\boldsymbol{n=}$ even.}
Consider the orbit closure $\clorbit{2^{n}}$ of $D_n\equiv D_{2l}$ with the 
corresponding Coulomb branch quiver
\begin{subequations}
\begin{align}
(2^n)^I: \qquad 
\raisebox{-.5\height}{
 	\begin{tikzpicture}
	\tikzstyle{gauge} = [circle, draw];
	\tikzstyle{flavour} = [regular polygon,regular polygon sides=4, draw];
	\node (g1) [gauge,label=below:{$1$}] {};
	\node (g2) [gauge, right of =g1,label=below:{$2$}] {};
	\node (g3) [ right of =g2] {$\ldots$};
	\node (g4) [gauge, right of =g3,label=below:{$2l{-}3$}] {};
	\node (g5) [gauge,right of =g4,label=below:{$2l{-}2$}] {};
	\node (g6) [gauge,above right of =g5,label=below:{$l$}] {};
	\node (g7) [gauge,below right of =g5,label=right:{$l{-}1$}] {};
	\node (f1) [flavour,right of=g6,label=below:{$2$}] {};
	\draw (g1)--(g2) (g2)--(g3) (g3)--(g4) (g4)--(g5) (g5)--(g6) (f1)--(g6) 
(g5)--(g7);
	\end{tikzpicture} \;,
	} \\
(2^n)^{II}: \qquad 
\raisebox{-.5\height}{
 	\begin{tikzpicture}
	\tikzstyle{gauge} = [circle, draw];
	\tikzstyle{flavour} = [regular polygon,regular polygon sides=4, draw];
	\node (g1) [gauge,label=below:{$1$}] {};
	\node (g2) [gauge, right of =g1,label=below:{$2$}] {};
	\node (g3) [ right of =g2] {$\ldots$};
	\node (g4) [gauge, right of =g3,label=below:{$2l{-}3$}] {};
	\node (g5) [gauge,right of =g4,label=below:{$2l{-}2$}] {};
	\node (g6) [gauge,below right of =g5,label=below:{$l$}] {};
	\node (g7) [gauge,above right of =g5,label=right:{$l{-}1$}] {};
	\node (f1) [flavour,right of=g6,label=below:{$2$}] {};
	\draw (g1)--(g2) (g2)--(g3) (g3)--(g4) (g4)--(g5) (g5)--(g6) (f1)--(g6) 
(g5)--(g7);
	\end{tikzpicture} \;.
	}
\end{align}
\end{subequations}
Note that one obtains two quiver gauge theories, because the very even 
partition $(2^n)$ corresponds to two nilpotent orbits.
The monopole formula admits non-trivial resolution parameters as there is a 
$\urm(2)$ flavour node present. Based on the examples computed, we expect the 
HWG to be
\begin{subequations}
\begin{align}
(2^n)^{I}: \qquad 
 \HWG_{(p_{1},p_{2})}^{(2^{n})} (t^2) &=
 x_{n-1}^{p_{1}+p_{2}}
     \left(\mu_{n-1} t^{l} \right)^{p_{1}-p_{2}} 
     \cdot 
     \PE\left[\sum_{i=1}^{l-1} \mu_{2i} t^{2i} + \mu_{n-1}^2 t^{2l} \right] 
     \;, \\
(2^n)^{II}: \qquad 
 \HWG_{(p_{1},p_{2})}^{(2^{n})} (t^2) &=
 x_{n}^{p_{1}+p_{2}}
     \left(\mu_{n} t^{l} \right)^{p_{1}-p_{2}} 
     \cdot 
     \PE\left[\sum_{i=1}^{l-1} \mu_{2i} t^{2i}  + \mu_n^2 
t^{2l} \right] \; .
\end{align}
\end{subequations}
Comparing this to the known results of \cite{Hesselink:1978}, the symplectic 
resolution is of the form
\begin{align}
 \pi_{(2^{n})} : T^* \left( \tfrac{\sorm(2n)}{\surm(n)\times\uo} \right) \to 
\clorbit{(2^{n})}
\end{align}
The HWG indicates this resolution behaviour due to the prefactors $\mu_{n-1}$ 
or $\mu_n$ such that the corresponding HSS are the semi-simple orbits through 
$\mu_{n-1}$ or $\mu_n$, respectively.

\paragraph{Remarks.}
Two remarks are in order: firstly, the accidental isomorphism $D_3\cong A_3$ is 
manifest in the Hilbert series results. To see this, compare $A$-type $(2^2)$ 
of Table \ref{tab:results_A-type_examples} with $D$-type $(3,1^3)$ of Table 
\ref{tab:results_D-type_examples}, or compare $A$-type $(2,1^2)$ with $D$-type 
$(2^2,1^2)$. In both cases, the results agree upon identification of fugacity 
and weight labels.

Secondly, for $\sorm(8)$ there is a triality relating the Coulomb branch 
quivers $(3,1^5)$ and $(2^4)^{I\slash II}$ via outer automorphism on $D_4$, 
see for instance Table \ref{tab:results_D-type_examples}. The triality rotates 
the fugacity and weight label of the node the flavour is attached to.

% 
%%%%%%%%%%%%%%%%%%%%%%%%%%%%%%%%%%%%%%%%%%%%%%%%%%%%%%%%%%%%%%%%%%%%%%%%%%%%%%%%
  \section{Exceptional algebras}
\label{sec:E-type}
Lastly, we consider the nilpotent orbits of the exceptional algebras, in 
particularly focusing on the characteristic height two examples of 
\cite{Hanany:2017ooe}.
The Hilbert series and HWG for the singular Coulomb branch have been presented 
in \cite{Hanany:2017ooe} and, here, we compute the monopole formula in the 
presence of background charges. We summarise the results in Tables 
\ref{tab:results_E-type_examples_1} and \ref{tab:results_E-type_examples_2}.

\begin{table}[!ht]
\centering
 \begin{tabular}{c|c|c|c}
 \toprule
  characteristic & $\dim_\C$  & quiver & HWG with flux\\ \midrule
  $\{1,0\}$  & $6 $ &
 	\raisebox{-.5\height}{
 	\begin{tikzpicture}
	\tikzset{node distance = 0.5cm}
	\tikzstyle{gauge} = [circle, draw,inner sep=2.5pt];
	\tikzstyle{flavour} = [regular polygon,regular polygon sides=4,inner 
sep=2.5pt, draw];
	\node (g1) [gauge,label=below:{$2$}] {};
	\node (g0) [right of =g1,xshift=-0.02cm] {};
	\node (g2) [gauge,right of =g0,xshift=-0.48cm,label=below:{$1$}] {};
	\node (f1) [flavour,above of=g1,label=above:{$1$}] {};
	\draw (g1)--(f1);
	\triplearrow{arrows={}}{(g1) -- (g2)};
	\triplearrow{arrows={-Implies}}{(g1) -- (g0)};	
	\end{tikzpicture}
	}
& no resolution
  \\
\midrule
$\{1,0,0,0\}$ & $16 $ &
\raisebox{-.5\height}{
	\begin{tikzpicture}
	\tikzset{node distance = 0.5cm}
	\tikzstyle{gauge} = [circle, draw,inner sep=2.5pt];
	\tikzstyle{flavour} = [regular polygon,regular polygon sides=4,inner 
sep=2.5pt, draw];
	\node (g1) [gauge,label=below:{$2$}] {};
	\node (g2) [gauge,right of =g1,label=below:{$3$}] {};
	\node (g0) [right of =g2,xshift=-0.05cm] {};
	\node (g3) [gauge,right of =g0,xshift=-0.45cm,label=below:{$2$}] {};
	\node (g4) [gauge,right of =g3,label=below:{$1$}] {};
	\node (f1) [flavour,above of=g1,label=above:{$1$}] {};
	\draw (g1)--(f1) (g1)--(g2) (g3)--(g4);
	\doublearrow{arrows={}}{(g2) -- (g3)};
	\doublearrow{arrows={-Implies}}{(g2) -- (g0)};	
	\end{tikzpicture}
	}
& no resolution \\ 
$\{0,0,0,1\}$ & $22 $ &
\raisebox{-.5\height}{
\begin{tikzpicture}
	\tikzset{node distance = 0.5cm}
	\tikzstyle{gauge} = [circle, draw,inner sep=2.5pt];
	\tikzstyle{flavour} = [regular polygon,regular polygon sides=4,inner 
sep=2.5pt, draw];
	\node (g1) [gauge,label=below:{$2$}] {};
	\node (g2) [gauge,right of =g1,label=below:{$4$}] {};
	\node (g0) [right of =g2,xshift=-0.05cm] {};
	\node (g3) [gauge,right of =g0,xshift=-0.45cm,label=below:{$3$}] {};
	\node (g4) [gauge,right of =g3,label=below:{$2$}] {};
	\node (f1) [flavour,above of=g4,label=above:{$1$}] {};
	\draw (g4)--(f1) (g1)--(g2) (g3)--(g4);
	\doublearrow{arrows={}}{(g2) -- (g3)};
	\doublearrow{arrows={-Implies}}{(g2) -- (g0)};	
	\end{tikzpicture}
      }
& no resolution \\
\bottomrule
 \end{tabular}
\caption{Coulomb branch quiver gauge theories for the exceptional algebras 
$G_2$ and $F_4$: realisations of the (closures of the) nilpotent orbits of 
characteristic height 2.}
\label{tab:results_E-type_examples_1}
\end{table}

\begin{table}[!ht]
\centering
 \begin{tabular}{c|c|c|c}
 \toprule
  characteristic &$\dim_\C$  &  quiver & HWG with flux \\ \midrule
 $\{0,0,0,0,0,1\}$ & $22 $ &
 \raisebox{-.5\height}{
 \begin{tikzpicture}
	\tikzset{node distance = 0.5cm}
	\tikzstyle{gauge} = [circle, draw,inner sep=2.5pt];
	\tikzstyle{flavour} = [regular polygon,regular polygon sides=4,inner 
sep=2.5pt, draw];
	\node (g1) [gauge,label=below:{$1$}] {};
	\node (g2) [gauge,right of =g1,label=below:{$2$}] {};
	\node (g3) [gauge,right of =g2,label=below:{$3$}] {};
	\node (g4) [gauge,right of =g3,label=below:{$2$}] {};
	\node (g5) [gauge,right of =g4,label=below:{$1$}] {};
	\node (g6) [gauge,above of =g3,label=right:{$2$}] {};
	\node (f1) [flavour,above of=g6,label=above:{$1$}] {};
	\draw (g1)--(g2) (g2)--(g3) (g3)--(g4) (g4)--(g5) (g3)--(g6) (g6)--(f1);
	\end{tikzpicture}
	}
 & no resolution \\ 
 $\{1,0,0,0,1,0\}$ &  $32 $ &
\raisebox{-.5\height}{
\begin{tikzpicture}
	\tikzset{node distance = 0.5cm}
	\tikzstyle{gauge} = [circle, draw,inner sep=2.5pt];
	\tikzstyle{flavour} = [regular polygon,regular polygon sides=4,inner 
sep=2.5pt, draw];
	\node (g1) [gauge,label=below:{$2$}] {};
	\node (g2) [gauge,right of =g1,label=below:{$3$}] {};
	\node (g3) [gauge,right of =g2,label=below:{$4$}] {};
	\node (g4) [gauge,right of =g3,label=below:{$3$}] {};
	\node (g5) [gauge,right of =g4,label=below:{$2$}] {};
	\node (g6) [gauge,above of =g3,label=above:{$2$}] {};
	\node (f1) [flavour,above of=g1,label=above:{$1$}] {};
	\node (f2) [flavour,above of=g5,label=above:{$1$}] {};
	\draw (g1)--(g2) (g2)--(g3) (g3)--(g4) (g4)--(g5) (g3)--(g6) (g1)--(f1) 
(g5)--(f2);
	\end{tikzpicture}
      }
& 
$ 
 \begin{cases}
 x_1^{p_1} x_5^{p_5}  (\mu_5 t^2)^{p_1 -p_5} \PE[\mu_6 t^2 {+} \mu_1 \mu_5 t^4] 
&, 
p_1 {\geq} p_5 \\
  x_1^{p_1} x_5^{p_5} (\mu_1 t^2)^{p_5 -p_1} \PE[\mu_6 t^2 {+} \mu_1 \mu_5 t^4] 
&, 
p_1 {\leq} p_5
 \end{cases}$
\\
\midrule 
$\{1,0,0,0,0,0,0\}$ &  $34 $ &
\raisebox{-.5\height}{
\begin{tikzpicture}
	\tikzset{node distance = 0.5cm}
	\tikzstyle{gauge} = [circle, draw,inner sep=2.5pt];
	\tikzstyle{flavour} = [regular polygon,regular polygon sides=4,inner 
sep=2.5pt, draw];
	\node (g1) [gauge,label=below:{$2$}] {};
	\node (g2) [gauge,right of =g1,label=below:{$3$}] {};
	\node (g3) [gauge,right of =g2,label=below:{$4$}] {};
	\node (g4) [gauge,right of =g3,label=below:{$3$}] {};
	\node (g5) [gauge,right of =g4,label=below:{$2$}] {};
	\node (g6) [gauge,right of =g5,label=below:{$1$}] {};
	\node (g7) [gauge,above of =g3,label=above:{$2$}] {};
	\node (f1) [flavour,above of=g1,label=above:{$1$}] {};
	\draw (g1)--(g2) (g2)--(g3) (g3)--(g4) (g4)--(g5) (g5)--(g6) (g3)--(g7) 
(g1)--(f1);
	\end{tikzpicture}
      }
& no resolution \\ 
$\{0,0,0,0,1,0,0\}$ &  $52 $ &
\raisebox{-.5\height}{
\begin{tikzpicture}
	\tikzset{node distance = 0.5cm}
	\tikzstyle{gauge} = [circle, draw,inner sep=2.5pt];
	\tikzstyle{flavour} = [regular polygon,regular polygon sides=4,inner 
sep=2.5pt, draw];
	\node (g1) [gauge,label=below:{$2$}] {};
	\node (g2) [gauge,right of =g1,label=below:{$4$}] {};
	\node (g3) [gauge,right of =g2,label=below:{$6$}] {};
	\node (g4) [gauge,right of =g3,label=below:{$5$}] {};
	\node (g5) [gauge,right of =g4,label=below:{$4$}] {};
	\node (g6) [gauge,right of =g5,label=below:{$2$}] {};
	\node (g7) [gauge,above of =g3,label=above:{$3$}] {};
	\node (f1) [flavour,above of=g5,label=above:{$1$}] {};
	\draw (g1)--(g2) (g2)--(g3) (g3)--(g4) (g4)--(g5) (g5)--(g6) (g3)--(g7) 
(g5)--(f1);
	\end{tikzpicture}
    }
& no resolution \\ 
$\{0,0,0,0,0,2,0\}$ &  $54 $ &
\raisebox{-.5\height}{
\begin{tikzpicture}
	\tikzset{node distance = 0.5cm}
	\tikzstyle{gauge} = [circle, draw,inner sep=2.5pt];
	\tikzstyle{flavour} = [regular polygon,regular polygon sides=4,inner 
sep=2.5pt, draw];
	\node (g1) [gauge,label=below:{$2$}] {};
	\node (g2) [gauge,right of =g1,label=below:{$4$}] {};
	\node (g3) [gauge,right of =g2,label=below:{$6$}] {};
	\node (g4) [gauge,right of =g3,label=below:{$5$}] {};
	\node (g5) [gauge,right of =g4,label=below:{$4$}] {};
	\node (g6) [gauge,right of =g5,label=below:{$3$}] {};
	\node (g7) [gauge,above of =g3,label=above:{$3$}] {};
	\node (f1) [flavour,above of=g6,label=above:{$2$}] {};
	\draw (g1)--(g2) (g2)--(g3) (g3)--(g4) (g4)--(g5) (g5)--(g6) (g3)--(g7) 
(g6)--(f1);
	\end{tikzpicture}
      }
& 
$x_6^{p_1+p_2} (\mu_6 t^{3})^{p_1 -p_2}\cdot 
 \PE[\mu_1 t^2 {+} \mu_5 t^4 {+} \mu_6^2 t^6]
$
\\
\midrule 
$\{0,0,0,0,0,0,1,0\}$ & $58 $ &
\raisebox{-.5\height}{
	\begin{tikzpicture}
	\tikzset{node distance = 0.5cm}
	\tikzstyle{gauge} = [circle, draw,inner sep=2.5pt];
	\tikzstyle{flavour} = [regular polygon,regular polygon sides=4,inner 
sep=2.5pt, draw];
	\node (g1) [gauge,label=below:{$2$}] {};
	\node (g2) [gauge,right of =g1,label=below:{$4$}] {};
	\node (g3) [gauge,right of =g2,label=below:{$6$}] {};
	\node (g4) [gauge,right of =g3,label=below:{$5$}] {};
	\node (g5) [gauge,right of =g4,label=below:{$4$}] {};
	\node (g6) [gauge,right of =g5,label=below:{$3$}] {};
	\node (g7) [gauge,right of =g6,label=below:{$2$}] {};
	\node (g8) [gauge,above of =g3,label=above:{$3$}] {};
	\node (f1) [flavour,above of=g7,label=above:{$1$}] {};
	\draw (g1)--(g2) (g2)--(g3) (g3)--(g4) (g4)--(g5) (g5)--(g6) (g6)--(g7) 
(g3)--(g8)
(g7)--(f1);
	\end{tikzpicture}
	}
& no resolution \\ 
$\{1,0,0,0,0,0,0,0\}$ & $92 $ &
\raisebox{-.5\height}{
\begin{tikzpicture}
	\tikzset{node distance = 0.5cm}
	\tikzstyle{gauge} = [circle, draw,inner sep=2.5pt];
	\tikzstyle{flavour} = [regular polygon,regular polygon sides=4,inner 
sep=2.5pt, draw];
	\node (g1) [gauge,label=below:{$4$}] {};
	\node (g2) [gauge,right of =g1,label=below:{$7$}] {};
	\node (g3) [gauge,right of =g2,label=below:{$10$}] {};
	\node (g4) [gauge,right of =g3,label=below:{$8$}] {};
	\node (g5) [gauge,right of =g4,label=below:{$6$}] {};
	\node (g6) [gauge,right of =g5,label=below:{$4$}] {};
	\node (g7) [gauge,right of =g6,label=below:{$2$}] {};
	\node (g8) [gauge,above of =g3,label=above:{$5$}] {};
	\node (f1) [flavour,above of=g1,label=above:{$1$}] {};
	\draw (g1)--(g2) (g2)--(g3) (g3)--(g4) (g4)--(g5) (g5)--(g6) (g6)--(g7) 
(g3)--(g8)
(g1)--(f1);
	\end{tikzpicture}
	}
	& no resolution \\
\bottomrule
 \end{tabular}
\caption{Coulomb branch quiver gauge theories for the exceptional algebras 
$E_6$, $E_7$, $E_8$: realisations of the (closures of the) nilpotent orbits of 
characteristic height two. 
The unphysical $\uo$ in $G_F$ can be eliminated as before: 
introducing an auxiliary fugacity $z_0$ for $\uo \subset G_F$, and imposing 
$z_0^{\sum_j p_j} \prod_{i=1}^n z_i^{r_i}=1$, with $r_i$ the ranks of the 
gauge nodes, leads to $x_1 x_5=1$ for $\{1,0,0,0,1,0\}$ and $x_6=1$ for 
$\{0,0,0,0,0,2,0\}$ (and $z_0\equiv 1$).}
\label{tab:results_E-type_examples_2}
\end{table}

\paragraph{$\boldsymbol{G_2}$ and $\boldsymbol{F_4}$.}
The quiver gauge theories of Table \ref{tab:results_E-type_examples_1} exhibit 
only a single $\uo$ flavour nodes such that there is no resolution parameter 
available on the Coulomb branch. This is consistent with the literature, see 
for instance \cite[Proposition 3.21]{Fu:2003a}.

As a remark, there exists a Coulomb branch realisation
\begin{align}
  \raisebox{-.5\height}{
 	\begin{tikzpicture}
	\tikzstyle{gauge} = [circle, draw];
	\tikzstyle{flavour} = [regular polygon,regular polygon sides=4, draw];
	\node (g1) [gauge,label=below:{$2$}] {};
	\node (g2) [gauge,right of =g1,label=below:{$3$}] {};
	\node (f1) [flavour,above of=g1,label=above:{$1$}] {};
	\draw (g1)--(f1) (g1)--(g2);
	\draw [-] (1.15,0.1) arc (168:0:16pt);
	\draw [-] (1.15,-0.1) arc (-168:0:16pt);
	\end{tikzpicture}
	}
\end{align}
for the $10$-dimensional nilpotent orbit $\orbit{\{2,0\}}$ of $G_2$ although it 
has height larger than two.
Although the $\uo$ flavour symmetry naively suggests that no symplectic 
resolution exists, it is known that the symplectic resolution is of the form 
$T^*\left(G_2 \slash \urm(2) \right) \to \clorbit{\{2,0\}}$, see also 
\cite{Fu:2003a}. This indicates that the Coulomb branch construction of 
nilpotent orbit closures $\clorbit{}$ with $\height{\orbit{}}>2$  is still an 
open issue for exceptional Lie algebras \cite{Hanany:2017ooe}.

\paragraph{E-type.}
Inspecting the $E$-series of Table \ref{tab:results_E-type_examples_2}, we 
observe that only $\{1,0,0,0,1,0\}$ of $E_6$ and $\{0,0,0,0,0,2,0\}$ of $E_7$ 
admit non-trivial background charges on the Coulomb branch. All other Coulomb 
branches do not admit a resolution parameter, which is consistent with 
\cite[Proposition 3.21]{Fu:2003a}. 

Let us study $\{1,0,0,0,1,0\}$ of $E_6$ in more detail: again, we find two 
different 
behaviours of the HWG 
\begin{align}
  \HWG_{(p_1,p_5)}^{\{1,0,0,0,1,0\}} (t^2) = \begin{cases}
x_1^{p_1} x_5^{p_5}
    \left(\mu_5 t^2 \right)^{p_1 -p_5} 
    \cdot \PE\left[\mu_6 t^2 + \mu_1 \mu_5 t^4 \right] &, p_1 \geq p_5 \\
x_1^{p_1} x_5^{p_5}
   \left(\mu_1 t^2 \right)^{p_5 -p_1} 
   \cdot \PE\left[\mu_6 t^2 + \mu_1 \mu_5 t^4 \right] &, p_1 \leq p_5 
 \end{cases}
 \label{eq:HWG_E6}
\end{align}
depending on the relative sign between the two $\uo$ flavour charges. As 
before, one can eliminate the overall unphysical $\uo$ in $G_F$ via $p_1+p_5=0$ 
and reduce to a single (positive) resolution parameter $\propto \pm(p_1 -p_5)$.
The two different cases in \eqref{eq:HWG_E6} are a manifestation of the Mukai 
flop of $E$-type, see Table \ref{tab:Mukai_Dynkin}. 
Moreover, the corresponding resolution is given by 
\begin{align}
 \pi_{\{1,0,0,0,1,0\}}: T^* \left(\tfrac{E_6}{\sorm(10)\times \uo}\right) \to 
\clorbit{\{1,0,0,0,1,0\}} \; .
\end{align}
We interpret the prefactor $\propto \mu_1, \mu_5$ as indicating the 
exceptional fibre $\pi_{\{1,0,0,0,1,0\}}^{-1}(0)\cong 
\tfrac{E_6}{\sorm(10)\times \uo}$, i.e.\ the resolution is given by the 
cotangent bundle of the HSS $E_{III}$ of Table \ref{tab:HSS}.

Next, we consider $\{0,0,0,0,0,2,0\}$ of $E_7$: here, the Coulomb branch 
computation yields
\begin{align}
 \HWG_{(p_1,p_2)}^{\{0,0,0,0,0,2,0\}} (t^2)=
 x_6^{p_1+p_2}
 \left(\mu_6 t^{3} \right)^{p_1 -p_2}
 \cdot  \PE\left[\mu_1 t^2 + \mu_5 t^4 + \mu_6^2 t^6\right] \;, 
 \label{eq:HWG_E7}
\end{align}
where one can eliminate the unphysical $\uo$ via $p_1+p_2=0$. Hence, one can 
reduce to a single 
$p_1-p_2$ background charge of $\surm(2)$, which gives the effective resolution 
parameter.
The HWG \eqref{eq:HWG_E7} is consistent with the expected resolution
\begin{align}
 \pi_{\{0,0,0,0,0,2,0\}}: T^*  \left(\tfrac{E_7}{E_6\times \uo} \right)\to 
\clorbit{\{0,0,0,0,0,2,0\}} \; .
\end{align}
We note in particular that there exists only one resolution and that the 
prefactor $\propto \mu_6$ suggests to be interpreted as manifestation of the 
exceptional fibre $\pi_{\{0,0,0,0,0,2,0\}}^{-1}(0)\cong \tfrac{E_7}{E_6\times 
\uo}$. This follows from the observation that the sixth fundamental weight of 
$E_7$ has stabiliser $E_6 \times \uo$ in $E_7$.

Therefore, all nilpotent orbits of characteristic height two that allow for a 
symplectic resolution are resolved by cotangent bundles of the Hermitian 
symmetric spaces $E_{III}$ and $E_{VII}$. The Coulomb branch construction via 
unitary 
quivers and its resolution via the monopole formula with background charges 
reproduce these features consistently.
\paragraph{Remark.}
So far, the resolutions of height two orbit closures have exhausted all 
Hermitian symmetric spaces, but has not produced all possible basic Mukai flops 
of Table \ref{tab:Mukai_Dynkin}. The missing piece would be the 
$50$-dimensional $E_6$ orbit of characteristic $\{0,1,0,1,0,0\}$, but it is not 
of characteristic height two such that no unitary quiver realisation is known 
\cite{Hanany:2017ooe}.  
% %%%%%%%%%%%%%%%%%%%%%%%%%%%%%%%%%%%%%%%%%%%%%%%%%%%%%%%%%%%%%%%%%%%%%%%%%%%%%%%%
  \section{Conclusions}
\label{sec:conclusions} 
In this paper we have examined to what extent the prescription of the monopole 
formula with background charges is suitable to study the resolutions of certain 
Coulomb branches, which are nilpotent orbit closures of (characteristic) height 
two. 
For the examples considered with $T^*(G\slash P) \to \clorbit{}$ such that 
$G\slash P \cong \orbit{\mu}^{\mathrm{ss}}$, the HWG takes a  
remarkably simple form
\begin{align}
 \HWG_{\mathrm{flux}} = \left(\mu \ t^\Delta \right)^{\mathrm{flux}} \cdot 
\HWG_{\mathrm{singular}} \,,
\label{eq:HWG_result}
\end{align}
which allows us to show that the monopole formula is consistent with 
the following known facts about resolutions of $\clorbit{}$:
\begin{compactenum}[(i)]
 \item Number of resolution parameters: If the flavour symmetry is a 
single $\uo$ then the flux associated can simply be absorbed by a redefinition 
of the GNO magnetic weights. Hence, the monopole formula does not admit any 
resolution parameter. If the flavour symmetry group is larger, there exists a 
non-trivial resolution. We note in particular, that an overall 
$\uo$ in $G_F$ is unphysical; for instance, the resolution parameter for 
$G_F{=}( \uo{\times} \uo)\slash \uo$ is either $p_1 -p_2\geq0$ or $p_2 
-p_1\geq0$, while 
for $G_F{=} \urm(2) \slash \uo$ the resolution parameter is $p_1- p_2\geq 0$.  
\item Form of resolution: the monopole formula results indicate that the 
exceptional fibre of the resolution can be read off from the prefactor in 
\eqref{eq:HWG_result}. In more detail, if $\clorbit{}$ is resolved by the 
cotangent bundle $T^*(G\slash P)$ of a HSS $G\slash P$, then the prefactor is 
proportional to a single fundamental weight $\mu$, such that the HSS is 
realised as semi-simple orbit $\orbit{\mu}^{\mathrm{ss}}$ through $\mu$ in 
$\Lie(G)$. Moreover, the size of the coset $G\slash P$, as base manifold of the 
resolved space, is determined by the exponent of $t^{\Delta \cdot 
\text{flux}}$.  
\item Existence of multiple resolutions, i.e.\ Mukai flops: the cases with two 
$\uo$ flavour nodes allow for two distinct cases, depending on which $\uo$ flux 
is larger. The HWG becomes case dependent and we interpret this as the 
manifestation of the basic Mukai flops. The examples considered reproduce all 
but one 
of the possible basic flops, see Table \ref{tab:Mukai_Dynkin}. The exception 
$E_{6,II}$ does not correspond to a height two nilpotent orbit.
\end{compactenum}
As a remark, existence of symplectic resolutions and, in case of multiple 
resolutions, their relation via (sequences of) Mukai flops can equivalently be 
discussed via weighted Dynkin diagrams. A brief summary is provided in Appendix 
\ref{app:weighted_Dynkin}.

Additionally, it is interesting to note that all Hermitian symmetric spaces of 
Table \ref{tab:HSS} are realised as base spaces for the resolutions of the 
(characteristic) height two nilpotent orbits. We summarise the orbits considered 
and their resolutions in Tables \ref{tab:summary_resolution_classical} -- 
\ref{tab:summary_resolution_exceptional}.
\begin{table}[t]
 \centering
 \begin{tabular}{c|c|c}
 \toprule
  type & partition & resolution \\ \midrule
$A_n$ & $\rho=(2^k,1^{n+1-2k})$, $2\leq 2k \leq n+1$ &  
$T^* \left(
\tfrac{ \surm(n+1) }{ \mathrm{S}(\urm(n+1-k)\times\urm(k)) }
\right)
\to \clorbit{\rho}$ 
\\ \midrule 
$B_n$ & $\rho=(2^{2k},1^{2n+1-4k})$, $k>0$ &
---
\\
 & $\rho=(3,1^{2n-2})$ &
$T^* \left(
\tfrac{ \sorm(2n+1) }{ \sorm(2n-1)\times\sorm(2) }
\right)
\to \clorbit{\rho}$  
\\ \midrule
$C_n$ & $\rho=(2^{k},1^{2(n-k)})$, $n>k\geq 1$ &
--- 
\\
 & $\rho=(2^n)$ &
$T^* \left(
\tfrac{ \sprm(n) }{ \urm(n) }
\right)
\to \clorbit{\rho}$  
\\ \midrule
$D_n$ & $\rho=(2^{2k},1^{2n-4k})$, $n-1>2k\geq 2$ &
---
\\
  & $\rho=(2^{n-1},1^{2})$, $n=$ odd &
  \multirow{2}{*}{$\bigg\}\quad$ $ T^* \left(
\tfrac{ \sorm(2n) }{ \urm(n) }
\right)
\to \clorbit{\rho}$ }
\\
 & $\rho=(2^{n})$, $n=$ even &
\\
 & $\rho=(3,1^{2n-3})$ &
$T^* \left(
\tfrac{ \sorm(2n) }{ \sorm(2n-2)\times \sorm(2) }
\right)
\to \clorbit{\rho}$  
\\ \bottomrule
\end{tabular}
\caption{Summary of resolutions of height two nilpotent orbits for classical 
algebras.}
\label{tab:summary_resolution_classical}
\end{table}

\begin{table}[t]
 \centering
 \begin{tabular}{c|c|c}
  \toprule
  type & characteristic & resolution \\ \midrule
$G_2$  & $\{1,0\}$ & --- \\
%   &  & --- \\
 \midrule
 %%%%%%%%%%%%%%%%%%%%%%%%%%%%%%%
 $F_4$ & $\{1,0,0,0\}$ & --- \\
  & $\{0,0,0,1\}$ & --- \\ 
  \midrule
 %%%%%%%%%%%%%%%%%%%%%%%%%%%%%%%
 $E_6$ & $\{0,0,0,0,0,1\}$ & --- \\
  & $\{1,0,0,0,1,0\}$ &  $T^* \left(\tfrac{E_6}{\sorm(10)\times \uo}\right)\to 
\clorbit{\{1,0,0,0,1,0\}}$ \\ 
  \midrule
 %%%%%%%%%%%%%%%%%%%%%%%%%%%%%%%
 $E_7$ & $\{1,0,0,0,0,0,0\}$ & --- \\
  & $\{0,0,0,0,1,0,0\}$ & --- \\
  & $\{0,0,0,0,0,2,0\}$ & $T^* \left(\tfrac{E_7}{E_6\times \uo}\right)\to 
\clorbit{\{0,0,0,0,0,2,0\}}$  \\ 
  \midrule
 %%%%%%%%%%%%%%%%%%%%%%%%%%%%%%%
 $E_8$ & $\{0,0,0,0,0,0,1,0\}$ & --- \\
  & $\{1,0,0,0,0,0,0,0\}$ & --- \\ 
  \midrule
 %%%%%%%%%%%%%%%%%%%%%%%%%%%%%%%
 \end{tabular}
\caption{Summary of resolutions of characteristic height two nilpotent orbits 
for exceptional algebras.}
\label{tab:summary_resolution_exceptional}
\end{table}

In summary, the monopole formula, in combination with the unitary Coulomb 
branch quiver realisations for the $BCD$-type and exceptional nilpotent orbit 
closures of (characteristic) height two, exhibits all features of the 
symplectic resolutions correctly. Thus, the monopole formula with background 
charges is not only suitable for gluing techniques as indicated in 
\cite{Cremonesi:2014uva}, but is in fact a tool to study the geometry of the 
resolved spaces. On the other hand, the inclusion of background fluxes opens a 
window to study the $3$-dimensional $\Ncal=4$ theories with discrete real 
masses. 
\paragraph{Outlook.}
Coulomb branches of nilpotent orbit closures with $\height{\orbit{}}\geq3$ have 
not been considered here. Let us mention some of the generalisations which are 
to be expected. The flavour symmetry groups may become larger such 
that more than one resolution parameter becomes relevant. In this case, the 
different orderings of the fluxes will all be related by a locally trivial 
family of Mukai flops. To put it differently, the basic Mukai flops will 
not be sufficient to relate the different symplectic resolutions of a given 
orbit, but a finite family of Mukai flops will do.

As example, consider $\clorbit{(3,2,1)}$ of $A_5$ with Coulomb branch quiver
\begin{align}
\raisebox{-.5\height}{
  	\begin{tikzpicture}
	\tikzstyle{gauge} = [circle, draw];
	\tikzstyle{flavour} = [regular polygon,regular polygon sides=4, draw];
	\node (g1) [gauge,label=below:{$1$}] {};
	\node (g2) [gauge, right of=g1,label=below:{$2$}] {};
	\node (g3) [gauge, right of=g2,label=below:{$3$}] {};
	\node (g4) [gauge, right of=g3,label=below:{$3$}] {};
	\node (g5) [gauge, right of=g4,label=below:{$2$}] {};
	\node (f1) [flavour,above of=g3,label=above:{$1$}] {};
	\node (f2) [flavour,above of=g4,label=above:{$1$}] {};
	\node (f3) [flavour,above of=g5,label=above:{$1$}] {};
	\draw (g1)--(g2)
			(g2)--(g3)
			(g3)--(g4)
			(g4)--(g5)			
			(g3)--(f1)
			(g4)--(f2)
			(g5)--(f3)		
		;
	\end{tikzpicture}
	} 
\end{align}
where the three $\uo$ fluxes $p_4$, $p_5$, $p_6$ have six ordered permutations 
$p_{\sigma(4)} \geq p_{\sigma(5)} \geq p_{\sigma(6)}$. From this we expect six 
different resolutions of $\clorbit{(3,2,1)}$. In fact, comparing this to 
\cite[Example 4.6]{Namikawa:2006} there exist six polarisations 
$P_{\sigma(4),\sigma(5),\sigma(6)}$. Each gives rise to a resolution 
$T^*(\surm(6)\slash P_{\sigma(4),\sigma(5),\sigma(6)}) \to \clorbit{(3,2,1)}$, 
and any two are linked by birational maps in the form of a locally trivial 
family of Mukai flops of type $A$.
However, we do not expect the HWG with fluxes to respect the appealing form 
\eqref{eq:HWG_result}, simply because the HWG without fluxes does not have a 
simple $\PE$ anymore.

In addition, in \cite[Section 3.1]{Hanany:2017ooe} a Hilbert series formula 
for normalisations of nilpotent orbit closures has been proposed. In view of 
our final result \eqref{eq:HWG_result}, the formula seems suited to deal with 
resolutions which do not come from a simple flag variety. This localisation 
approach might also be able to compute the $E_6$ orbit of characteristic 
$\{0,1,0,1,0,0\}$, which is expected to realise the last basic Mukai flop 
$E_{6,II}$. 
\paragraph{Acknowledgements.}
The authors would like to thank Santiago Cabrera, Rudolph Kalveks, Karin Baur, 
and Charles Beil for useful discussions.
We thank the Simons Center for Geometry and Physics, Stony Brook University  
for the hospitality and the 
partial support during the early stage of this work at the Simons Summer 
workshop 2017.
We thank the Galileo Galilei Institute for
Theoretical Physics for the hospitality and the INFN for partial support
during the intermediate stage of this work at the workshop ``Supersymmetric 
Quantum Field Theories in the Non-perturbative Regime'' in 2018.
A.H.\ thanks the Mathematical Physics group of the University of Vienna for 
hospitality during the final stage of this work.
A.H.\ is supported by STFC Consolidated Grant ST/J0003533/1, and EPSRC 
Programme 
Grant EP/K034456/1.
M.S.\ is supported by Austrian Science Fund (FWF) grant P28590. 
M.S.\ thanks the Faculty of Physics of the University of Vienna for travel 
support via the ``Jungwissenschaftsförderung''.
%%%%%%%%%%%%%%%%%%%%%%%%%%%%%%%%%%%%%%%%%%%%%%%%%%%%%%%%%%%%%%%%%%%%%%%%%%%%%%%
\appendix
\section{Conventions}
\label{app:conventions}
In this appendix, we provide the conventions used in all calculations. We note 
that all Coulomb branch quivers $T_\Gamma$ are balanced and the set of balanced 
nodes forms exactly the Dynkin diagram $\Gamma$ of the algebra $\gfrak$ to 
which the nilpotent orbit closure $\MCoulomb(T_\Gamma)\cong \clorbit{}$ 
belongs. 
As such, the quiver $T_\Gamma$ has exactly $r=\rank(G)$ many unitary nodes, 
each node $i=1,\ldots, r$ is associated with a topological symmetry group 
$\uo_i$, which we count by a fugacity $z_i$. The root space fugacities $z_i$ 
are transformed into the weight space fugacities $x_i$ by means of the Cartan 
matrix $A_{ij}$ of the algebra $\gfrak$; in detail,
\begin{align}
 z_i = \prod_{j=1}^r x_j^{A_{ij}} 
 \qquad \Leftrightarrow \quad
 x_i = \prod_{j=1}^r z_j^{(A^{-1})_{ij}} \; . 
 \label{eq:root_to_weight}
\end{align}
Labelling of fugacities $z_i$ (or likewise $x_i$) and background charges $p$ 
will follow the numbering of the nodes in the corresponding Dynkin diagram, see 
for instance \cite[Table IV]{Fuchs:1997jv}. As example
\begin{align}
\text{$\uo$-flavours:} \quad
 \raisebox{-.5\height}{
 	\begin{tikzpicture}
	\tikzset{node distance = 0.5cm}
	\tikzstyle{gauge} = [circle, draw,inner sep=2.5pt];
	\tikzstyle{flavour} = [regular polygon,regular polygon sides=4,inner 
sep=2.5pt, draw];
	\node (g1) [gauge,label=below:{$z_1$}] {};
	\node (g2) [gauge,right of =g1,label=below:{$z_2$}] {};
	\node (g3) [gauge,right of =g2,label=below:{$z_3$}] {};
	\node (f1) [flavour,above of=g1,label=above:{$p_1$}] {};
	\node (f3) [flavour,above of=g3,label=above:{$p_3$}] {};
% 	\draw (g1)--(f1);
	\draw (g1)--(g2) (g2)--(g3) (g1)--(f1) (g3)--(f3);
	\end{tikzpicture}
	}
	\qquad
	\text{or}
	\qquad
	\text{$\utwo$-flavour:} \quad
 \raisebox{-.5\height}{
 	\begin{tikzpicture}
	\tikzset{node distance = 0.5cm}
	\tikzstyle{gauge} = [circle, draw,inner sep=2.5pt];
	\tikzstyle{flavour} = [regular polygon,regular polygon sides=4,inner 
sep=2.5pt, draw];
	\node (g1) [gauge,label=below:{$z_1$}] {};
	\node (g2) [gauge,right of =g1,label=below:{$z_2$}] {};
	\node (g3) [gauge,right of =g2,label=below:{$z_3$}] {};
	\node (f2) [flavour,above of=g2,label=above:{$p_1,p_2$}] {};
% 	\node (f3) [flavour,above of=g3,label=above:{$p_3$}] {};
% 	\draw (g1)--(f1);
	\draw (g1)--(g2) (g2)--(g3) (g2)--(f2);
	\end{tikzpicture}
	} \; .
\end{align}

In the relevant quivers, the gauge nodes are linked by different types of 
hyper multiplets, which have the following contributions to the conformal 
dimension:
\begin{subequations}
\begin{alignat}{3}
&\raisebox{-.5\height}{
 	\begin{tikzpicture}
	\tikzset{node distance = 1cm}
	\tikzstyle{gauge} = [circle, draw,inner sep=2.5pt];
% 	\tikzstyle{flavour} = [regular polygon,regular polygon sides=4,inner 
% sep=2.5pt, draw];
	\node (g1) [gauge,label=below:{$k$}] {};
	\node (g2) [gauge,right of =g1,label=below:{$l$}] {};
% 	\node (f1) [flavour,above of=g1,label=above:{$1$}] {};
% 	\draw (g1)--(f1);
	\draw (g1)--(g2);
	\end{tikzpicture}
	}
	&
	\qquad &\leftrightarrow \qquad
	&
	\Delta_{\text{h-plet}}&= \frac{1}{2}\sum_{n=1}^{k}\sum_{m=1}^{l} 
|q_{1,n}-q_{2,m}| \; ,
	\\
% 	
%%%%%%%%%%%%%%%%%%%%%%%%%%%%%%%%%%%%%%%%%%%%%%%%%%%%%%%% 
% 
&\raisebox{-.5\height}{
 	\begin{tikzpicture}
	\tikzset{node distance = 0.5cm}
	\tikzstyle{gauge} = [circle, draw,inner sep=2.5pt];
	\node (g1) [gauge,label=below:{$k$}] {};
	\node (g0) [right of =g1,xshift=+0.25cm] {};
	\node (g2) [gauge,right of =g0,xshift=-0.25cm,label=below:{$l$}] {};
	\doublearrow{arrows={}}{(g1) -- (g2)}
	\doublearrow{arrows={-Implies}}{(g1) -- (g0)}	
	\end{tikzpicture}
	}
		&
	\qquad &\leftrightarrow \qquad
	&
	\Delta_{\text{h-plet}}&= \frac{1}{2}\sum_{n=1}^{k}\sum_{m=1}^{l} 
|2\cdot q_{1,n}-q_{2,m}| \; ,
\label{eq:non-simply_laced_2}
	\\
% 	
%%%%%%%%%%%%%%%%%%%%%%%%%%%%%%%%%%%%%%%%%%%%%%%%%%%%%%
%
&\raisebox{-.5\height}{
 	\begin{tikzpicture}
	\tikzset{node distance = 0.5cm}
	\tikzstyle{gauge} = [circle, draw,inner sep=2.5pt];
	\node (g1) [gauge,label=below:{$k$}] {};
	\node (g0) [right of =g1,xshift=+0.25cm] {};
	\node (g2) [gauge,right of =g0,xshift=-0.25cm,label=below:{$l$}] {};
	\triplearrow{arrows={}}{(g1) -- (g2)}
	\triplearrow{arrows={-Implies}}{(g1) -- (g0)}	
	\end{tikzpicture}
	}
		&
	\qquad &\leftrightarrow \qquad
	&
	\Delta_{\text{h-plet}}&= \frac{1}{2}\sum_{n=1}^{k}\sum_{m=1}^{l} 
|3\cdot q_{1,n}-q_{2,m}|\; .
\label{eq:non-simply_laced_3}
\end{alignat}
\end{subequations}
Here, the magnetic charges of the $i$-th node $\urm(k_i)$ are labelled as 
$q_{i,n}$ for $n=1,\ldots,k_i$. The non-simply laced links of 
\eqref{eq:non-simply_laced_2}--\eqref{eq:non-simply_laced_3} have been 
introduced in \cite[Equation 3.3]{Cremonesi:2014xha}.

When working with Hilbert series (HS) and Highest Weight Generating functions 
(HWG), important tools are the Plethystic Exponential (PE) and Plethystic 
Logarithm (PL), the inverse of the PE. For our purpose, it is sufficient to note
\begin{align}
 \PE\left[\sum_{i=1}^c f_i t^{a_i} - \sum_{j=1}^d g_j t^{b_j} \right] \equiv 
\frac{\prod_{j=1}^{d} (1-g_j t^{b_j})}{\prod_{i=1}^c(1- f_i t^{a_i})} \; ,
\end{align}
where $a_i,b_j \in \NN$ are exponents and $f_j,g_j$ are monomials in weight or 
root fugacities. For further details, we refer to \cite{Benvenuti:2006qr}.
\section{Weighted Dynkin diagram}
\label{app:weighted_Dynkin}
Another method of labelling nilpotent orbits of $\gfrak$ employs weighted 
Dynkin diagrams 
(WDD), which decorate the nodes of the Dynkin diagram for $\gfrak$ with labels 
$0,1,$ or $2$. We refer to \cite{Collingwood:1993} for details. In contrast to 
partitions, WDDs are applicable to nilpotent orbits of classical as well as 
exceptional Lie 
algebras. On the other hand, the set of $ABCD$-type partitions is one-to-one to 
all $ABCD$-type nilpotent orbits; whereas, the number $3^{\rank(G)}$ of 
possible WDDs is significantly larger than the set of nilpotent orbits. To our 
understanding, no algorithm exists that could predict which diagrams are 
realised. 

To provide a more intuitive picture, all unitary Coulomb branch realisations of 
height two nilpotent orbit closures are built from the corresponding WDD in the 
following way: The WDD provides the ranks of the flavour nodes and the condition 
for the quiver to be balanced determines the ranks of the gauge nodes uniquely.

Following \cite{Fu:2007}, one introduces the sets $\Theta_i$, which collect all 
nodes in the WDD with label $i=0,1,2$. Focusing on height two, a result by  
\cite[Lemma 3.2]{Fu:2007} states the following: $ABC$-type nilpotent orbit 
closures admit a symplectic resolution if and only if the number of elements in 
$\Theta_1$ is even. For $D$-type orbit closures, a symplectic resolution exists 
if and only if either the number of elements in $\Theta_1$ is even, or the 
number of elements in $\Theta_1$ is even and the two spinor nodes belong to 
$\Theta_1$.   

The weighted Dynkin diagram allows to deduce the \emph{standard parabolic 
subalgebra} associated to the nilpotent orbit, and the set of all parabolic 
subalgebras can be labelled by marked Dynkin diagrams. 
As $|\Theta_1|$ is necessarily even for orbit closures that can be resolved, 
one can subdivide the set into two sets of equal length and try to construct 
two different parabolic subalgebras. This leads to 
\cite[Theorem 3.3]{Fu:2007}: (i) two symplectic resolutions for $A$-type, (ii) 
one symplectic resolution for $BC$-type, and (iii) two symplectic resolutions 
for $D$-type if the two spinor nodes are contained in $\Theta_1$, otherwise 
there exists only one resolution.

\begin{table}
\centering
 \begin{tabular}{c|c|c}
  \toprule 
  partition & WDD & MDD \\ \midrule
  $(4)$ &  
   \raisebox{-.5\height}{
    	\begin{tikzpicture}
    	\tikzset{node distance = 0.5cm}
	\tikzstyle{gauge} = [circle, draw,inner sep=2.5pt];
	\tikzstyle{dark} = [circle,draw,inner sep=2.5pt,fill=black];
	\node (g1) [gauge,label=below:{$2$}] {};
	\node (g2) [gauge,right of =g1,label=below:{$2$}] {};
	\node (g3) [gauge,right of =g2,label=below:{$2$}] {};
	\draw (g1)--(g2) (g2)--(g3);
	\end{tikzpicture}
	} &
	 \raisebox{-.5\height}{
    	\begin{tikzpicture}
    	\tikzset{node distance = 0.5cm}
	\tikzstyle{gauge} = [circle, draw,inner sep=2.5pt];
	\tikzstyle{dark} = [circle,draw,inner sep=2.5pt,fill=black];
	\node (g1) [dark] {};
	\node (g2) [dark,right of =g1] {};
	\node (g3) [dark,right of =g2] {};
	\draw (g1)--(g2) (g2)--(g3);
	\end{tikzpicture}
	} \\ 
  $(3,1)$ &  
   \raisebox{-.5\height}{
    	\begin{tikzpicture}
    	\tikzset{node distance = 0.5cm}
	\tikzstyle{gauge} = [circle, draw,inner sep=2.5pt];
	\tikzstyle{dark} = [circle,draw,inner sep=2.5pt,fill=black];
	\node (g1) [gauge,label=below:{$2$}] {};
	\node (g2) [gauge,right of =g1,label=below:{$0$}] {};
	\node (g3) [gauge,right of =g2,label=below:{$2$}] {};
	\draw (g1)--(g2) (g2)--(g3);
	\end{tikzpicture}
	} &
	 \raisebox{-.5\height}{
    	\begin{tikzpicture}
    	\tikzset{node distance = 0.5cm}
	\tikzstyle{gauge} = [circle, draw,inner sep=2.5pt];
	\tikzstyle{dark} = [circle,draw,inner sep=2.5pt,fill=black];
	\node (g1) [gauge] {};
	\node (g2) [dark,right of =g1] {};
	\node (g3) [dark,right of =g2] {};
	\draw (g1)--(g2) (g2)--(g3);
	\end{tikzpicture}
	} 
	or
	 \raisebox{-.5\height}{
    	\begin{tikzpicture}
    	\tikzset{node distance = 0.5cm}
	\tikzstyle{gauge} = [circle, draw,inner sep=2.5pt];
	\tikzstyle{dark} = [circle,draw,inner sep=2.5pt,fill=black];
	\node (g1) [dark] {};
	\node (g2) [gauge,right of =g1] {};
	\node (g3) [dark,right of =g2] {};
	\draw (g1)--(g2) (g2)--(g3);
	\end{tikzpicture}
	}
	or
	 \raisebox{-.5\height}{
    	\begin{tikzpicture}
    	\tikzset{node distance = 0.5cm}
	\tikzstyle{gauge} = [circle, draw,inner sep=2.5pt];
	\tikzstyle{dark} = [circle,draw,inner sep=2.5pt,fill=black];
	\node (g1) [dark] {};
	\node (g2) [dark,right of =g1] {};
	\node (g3) [gauge,right of =g2] {};
	\draw (g1)--(g2) (g2)--(g3);
	\end{tikzpicture}
	}
	\\ 
$(2^2)$ &  
   \raisebox{-.5\height}{
    	\begin{tikzpicture}
    	\tikzset{node distance = 0.5cm}
	\tikzstyle{gauge} = [circle, draw,inner sep=2.5pt];
	\tikzstyle{dark} = [circle,draw,inner sep=2.5pt,fill=black];
	\node (g1) [gauge,label=below:{$0$}] {};
	\node (g2) [gauge,right of =g1,label=below:{$2$}] {};
	\node (g3) [gauge,right of =g2,label=below:{$0$}] {};
	\draw (g1)--(g2) (g2)--(g3);
	\end{tikzpicture}
	} &
	 \raisebox{-.5\height}{
    	\begin{tikzpicture}
    	\tikzset{node distance = 0.5cm}
	\tikzstyle{gauge} = [circle, draw,inner sep=2.5pt];
	\tikzstyle{dark} = [circle,draw,inner sep=2.5pt,fill=black];
	\node (g1) [gauge] {};
	\node (g2) [dark,right of =g1] {};
	\node (g3) [gauge,right of =g2] {};
	\draw (g1)--(g2) (g2)--(g3);
	\end{tikzpicture}
	}
	\\
 $(2,1^2)$ &  
   \raisebox{-.5\height}{
    	\begin{tikzpicture}
    	\tikzset{node distance = 0.5cm}
	\tikzstyle{gauge} = [circle, draw,inner sep=2.5pt];
	\tikzstyle{dark} = [circle,draw,inner sep=2.5pt,fill=black];
	\node (g1) [gauge,label=below:{$1$}] {};
	\node (g2) [gauge,right of =g1,label=below:{$0$}] {};
	\node (g3) [gauge,right of =g2,label=below:{$1$}] {};
	\draw (g1)--(g2) (g2)--(g3);
	\end{tikzpicture}
	} &
	 \raisebox{-.5\height}{
    	\begin{tikzpicture}
    	\tikzset{node distance = 0.5cm}
	\tikzstyle{gauge} = [circle, draw,inner sep=2.5pt];
	\tikzstyle{dark} = [circle,draw,inner sep=2.5pt,fill=black];
	\node (g1) [dark] {};
	\node (g2) [gauge,right of =g1] {};
	\node (g3) [gauge,right of =g2] {};
	\draw (g1)--(g2) (g2)--(g3);
	\end{tikzpicture}
	}
	or
	\raisebox{-.5\height}{
    	\begin{tikzpicture}
    	\tikzset{node distance = 0.5cm}
	\tikzstyle{gauge} = [circle, draw,inner sep=2.5pt];
	\tikzstyle{dark} = [circle,draw,inner sep=2.5pt,fill=black];
	\node (g1) [gauge] {};
	\node (g2) [gauge,right of =g1] {};
	\node (g3) [dark,right of =g2] {};
	\draw (g1)--(g2) (g2)--(g3);
	\end{tikzpicture}
	}
	\\
 $(1^4)$ &  
   \raisebox{-.5\height}{
    	\begin{tikzpicture}
    	\tikzset{node distance = 0.5cm}
	\tikzstyle{gauge} = [circle, draw,inner sep=2.5pt];
	\tikzstyle{dark} = [circle,draw,inner sep=2.5pt,fill=black];
	\node (g1) [gauge,label=below:{$0$}] {};
	\node (g2) [gauge,right of =g1,label=below:{$0$}] {};
	\node (g3) [gauge,right of =g2,label=below:{$0$}] {};
	\draw (g1)--(g2) (g2)--(g3);
	\end{tikzpicture}
	} &
	 \raisebox{-.5\height}{
    	\begin{tikzpicture}
    	\tikzset{node distance = 0.5cm}
	\tikzstyle{gauge} = [circle, draw,inner sep=2.5pt];
	\tikzstyle{dark} = [circle,draw,inner sep=2.5pt,fill=black];
	\node (g1) [gauge] {};
	\node (g2) [gauge,right of =g1] {};
	\node (g3) [gauge,right of =g2] {};
	\draw (g1)--(g2) (g2)--(g3);
	\end{tikzpicture}
	}
	\\ \bottomrule
 \end{tabular}
\caption{Nilpotent orbits of $A_3$ can be labelled by partitions or weighted 
Dynkin diagrams. The corresponding parabolic subalgebras can be encoded in 
marked Dynkin diagrams.}
\label{tab:weighted_Dynkin}
\end{table}

As an example, the nilpotent orbits of $A_3$, labelled by their 
weighted Dynkin diagrams, and the corresponding parabolic subalgebras, labelled 
by marked Dynkin diagrams (MDD), are summarised in Table 
\ref{tab:weighted_Dynkin}. All (non-trivial) orbits of $A_3$ admit symplectic 
resolutions, as there are either zero nodes with weight label $1$ or exactly 
two. The existence of multiple resolutions for a single orbit closure can be 
observed from the marked Dynkin diagrams. A way to relate all possible 
parabolic subalgebras that give rise to different resolutions of the same orbit 
closure is presented in \cite[Definition 1]{Namikawa:2004} by means of 
operations on the MDD. Hence, the two (dual) MDDs of $(2,1^2)$ are the basic 
$A$-type Mukai flop, while the three MDDs of $(3,1)$ can be obtain by locally 
applying the $A$-type Mukai flop of $A_2$, i.e. the dual MDDs are
\raisebox{-.0\height}{
    	\begin{tikzpicture}
    	\tikzset{node distance = 0.5cm}
	\tikzstyle{gauge} = [circle, draw,inner sep=2.5pt];
	\tikzstyle{dark} = [circle,draw,inner sep=2.5pt,fill=black];
	\node (g1) [dark] {};
	\node (g2) [gauge,right of =g1] {};
	\draw (g1)--(g2);
	\end{tikzpicture}
	}
and 
	\raisebox{-.0\height}{
    	\begin{tikzpicture}
    	\tikzset{node distance = 0.5cm}
	\tikzstyle{gauge} = [circle, draw,inner sep=2.5pt];
	\tikzstyle{dark} = [circle,draw,inner sep=2.5pt,fill=black];
	\node (g1) [gauge] {};
	\node (g2) [dark,right of =g1] {};
	\draw (g1)--(g2);
	\end{tikzpicture}
	}.
Therefore, $\clorbit{(3,1)}$ of $A_3$ admits three different resolutions, which 
can also be seen from the corresponding Coulomb branch quiver
\begin{align}
\raisebox{-.5\height}{
  	\begin{tikzpicture}
	\tikzstyle{gauge} = [circle, draw];
	\tikzstyle{flavour} = [regular polygon,regular polygon sides=4, draw];
	\node (g1) [gauge,label=below:{$1$}] {};
	\node (g2) [gauge, right of=g1,label=below:{$2$}] {};
	\node (g3) [gauge, right of=g2,label=below:{$2$}] {};
	\node (f2) [flavour,above of=g2,label=above:{$1$}] {};
	\node (f3) [flavour,above of=g3,label=above:{$2$}] {};
	\draw (g1)--(g2)
			(g2)--(g3)			
			(g2)--(f2)
			(g3)--(f3)		
		;
	\end{tikzpicture}
	} 
\end{align}
in which one would identify the three resolutions by the relative size of the 
$\uo$ flux $p_1$ with respect to the $\urm(2)$ fluxes $p_2\geq p_3$.
%%%%%%%%%%%%%%%%%%%%%%%%%%%%%%%%%%%%%%%%%%%%%%%%%%%%%%%%%%%%%%%%%%%%%%%%%%
% 
 \bibliographystyle{JHEP}     %
 {\footnotesize{\bibliography{references}}}

\end{document}